\documentclass[12pt]{article}
\usepackage{a4wide}
\usepackage{amssymb, amsmath}
\usepackage{subcaption, graphicx}
\usepackage{cite}
\usepackage[utf8]{inputenc}
\usepackage[colorlinks]{hyperref}
\usepackage[usenames,dvipsnames]{color}
\hypersetup{
     breaklinks=true,
    pdfstartview={FitH},    % fits the width of the page to the window
    colorlinks=true,       % false: boxed links; true: colored links
    linkcolor=blue,          % color of internal links
    citecolor=red,        % color of links to bibliography
    filecolor=magenta,      % color of file links
    urlcolor=blue,           % color of external links
    anchorcolor=green,      % Color for anchor text
    linktocpage=true
}
\graphicspath{{figures/}}
\DeclareMathOperator{\sgn}{sgn}
\begin{document}
{\renewcommand{\thefootnote}{\fnsymbol{footnote}}
		
\begin{center}
{\LARGE Multi-field inflation from single-field models} 
\vspace{1.5em}

Martin Bojowald,$^{1}$\footnote{e-mail address: {\tt bojowald@gravity.psu.edu}},
Suddhasattwa Brahma,$^{2}$\footnote{e-mail address: {\tt
    suddhasattwa.brahma@gmail.com}},
Sean Crowe,$^3$\footnote{e-mail address: {\tt sean.crowe.92@gmail.com}}\\
Ding Ding$^{1}$\footnote{e-mail address: {\tt dud79@psu.edu}}
and Joseph McCracken$^4$\footnote{e-mail address: {\tt jm2264@cornell.edu}}
\\
\vspace{0.5em}
$^1$  Institute for Gravitation and the Cosmos, The Pennsylvania State
  University,\\ 104  Davey Lab, University Park, PA 16802, USA\\
$^2$ Department of Physics, McGill University,
Montr\'eal, QC H3A 2T8, Canada\\
$^3$   Institute of Theoretical Physics, Jagiellonian University, ul.\ {\L}ojasiewicza 11, 30-348 Krak\'{o}w, Poland\\
and Department of Physics, Georgia Southern University, Savannah, GA 31419
USA\\
$^4$ Department of Physics, Cornell University, Ithaca, NY 14853, USA
\vspace{1.5em}
\end{center}
}
	
\setcounter{footnote}{0}

\newcommand{\bea}{\begin{eqnarray}}
\newcommand{\eea}{\end{eqnarray}}
\renewcommand{\d}{{\mathrm{d}}}
\renewcommand{\[}{\left[}
\renewcommand{\]}{\right]}
\renewcommand{\(}{\left(}
\renewcommand{\)}{\right)}
\newcommand{\nn}{\nonumber}
\def\H{\mathrm{H}}
\def\V{\mathrm{V}}
\def\e{\mathrm{e}}
\def\be{\begin{equation}}
\def\ee{\end{equation}}

\def\al{\alpha}
\def\bet{\beta}
\def\gam{\gamma}
\def\om{\omega}
\def\Om{\Omega}
\def\sig{\sigma}
\def\Lam{\Lambda}
\def\lam{\lambda}
\def\ep{\epsilon}
\def\ups{\upsilon}
\def\vep{\varepsilon}
\def\S{\mathcal{S}}
\def\doi{http://doi.org}
\def\arxiv{http://arxiv.org/abs}
\def\d{\mathrm{d}}
\def\g{\mathrm{g}}
\def\m{\mathrm{m}}
\def\r{\mathrm{r}}

\begin{abstract}
  \noindent Quantization implies independent degrees of freedom that do not
  appear in the classical theory, given by fluctuations, correlations, and
  higher moments of a state. A systematic derivation of the resulting
  dynamical systems is presented here in a cosmological application for
  near-Gaussian states of a single-field inflation model. As a consequence,
  single-field Higgs inflation is made viable observationally by becoming a
  multi-field model with a specific potential for a fluctuation field
  interacting with the inflaton expectation value. Crucially, non-adiabatic
  methods of semiclassical quantum dynamics reveal important phases that can
  set suitable initial conditions for slow-roll inflation (in combination with
  the uncertainty relation), and then end inflation after the observationally
  preferred number of $e$-folds. New parameters in the interaction potential
  are derived from properties of the underlying background state,
  demonstrating how background non-Gaussianity can affect observational
  features of inflation or, conversely, how observations may be used to
  understand the quantum state of the inflaton.
\end{abstract}

\section{Introduction}
It is a natural requirement that self-consistent inflationary models should be
largely independent of the high energy quantum gravity theory, viewed in an
effective field theory framework. However, an exact decoupling of scales
relevant for inflation from high-energy modes can happen only if the
low-energy Lagrangian consists entirely of terms that are renormalizable using
Wilsonian effective actions. This condition restricts single-field models of
inflation to be of chaotic type with quartic potentials. 

If the inflationary action contains terms beyond mass-dimension four, then the
theory is liable to be affected by as yet unknown high-energy physics. In
fact, one even has to rely on ultraviolet physics in order to derive a
suitable higher-order form of the potential. In common single-field inflation,
this problem can rarely be avoided as the models preferred by observations
\cite{PlanckPot} depend crucially on non-renormalizable terms in the
potential, as for instance in Starobinsky inflation
\cite{Starobinsky}. Fundamentally, such terms have to be understood as
remnants in an effective description of some underlying theory of gravity and
matter, such as quantum gravity or string theory, but specific top-down
justifications of suitable forms of the potential are usually hard to come by.

Alternatively, if chaotic-type potentials, which have been ruled out by
data as single-field models, can somehow be resurrected, then the burden of
explaining these potentials does not have to fall on quantum
gravity. Motivated by this observation, we begin with a Higgs-inspired
classical potential,
\begin{eqnarray}\label{Cl}	
	V_{\rm cl}(\psi) =M^4 \(1-2\dfrac{\psi^2}{v^2} +
        \dfrac{\psi^4}{v^4}\) 	
\end{eqnarray}
with two parameters, $M$ and $v$, assumed to be positive. While the only known
scalar to have been discovered to date is the standard-model Higgs particle,
it is well-known that this type of a inflaton potential, by itself, is found
to be inconsistent with cosmological observations. To make matters worse, even
renormalization-group improvements do not suffice to make Higgs-like
potentials compatible with data \cite{Higgs_prob1, Higgs_prob2, Higgs_prob3}.
The only observationally consistent formulations proposed up until now have
been based on a scalar field non-minimally coupled to the Ricci scalar
\cite{HiggsNonMin,HiggsInflation}, modifying the kinetic term of the Higgs
field. Non-minimal coupling terms, however, mean that one is forced to modify
the nature of the standard model at high energies \cite{Modified_SM}, amongst
other issues \cite{Naturalness_Nonmin_Higgs}.

In the present work, we will preserve the simple nature of a minimally coupled
field with a quartic classical potential \eqref{Cl}. Applying a canonical
formalism of effective theory which, crucially, remains valid in non-adiabatic
regimes, the classical potential will be quantum extended to a two-field model
with a specific potential derived from (\ref{Cl}). The second field,
$\varphi$, represents quantum fluctuations of the background inflaton,
$\psi$. As such, it is subject to uncertainty relations that will be used to
obtain important lower bounds on its initial value. Initial evolution is then
non-adiabatic, but it automatically sets the stage for a long slow-roll phase
(in a so-called waterfall regime of the two-field model) that is consistent
with observational constraints. A final non-adiabatic phase automatically ends
inflation with just the right number of $e$-folds in a large region of the
parameter space.

Coefficients of the two-field potential are determined by the same two
parameters, $M$ and $v$, that appear in the single-field model
(\ref{Cl}). In addition, there are new coefficients derived from moments of
the inflaton state, such as parameters for non-Gaussianity of the background
state. In inflation models, this is a new kind of non-Gaussianity different
from what one usually refers to in primordial fluctuations during
inflation. In our case, non-Gaussianity is present already in the wave
function of the homogeneous quantum inflaton field (referred to here as the
background state), and not only in the perturbation spectrum. It is therefore
possible to put constraints on the two-field potential based on known
properties of states, or conversely, to determine conditions on suitable
inflaton states based on observational constraints. An important finding is
that constraints on the spectral index, its running, and the tensor-to-scalar
ratio prefer small background non-Gaussianity.

In Section~\ref{s:EffPot}, we present a review of relevant methods of
non-adiabatic quantum dynamics, which have appeared in various forms in fields
as diverse as quantum field theory, quantum chaos, quantum chemistry, and
quantum cosmology. The same section presents a comparison with Gaussian
methods and shows how non-adiabatic dynamics can include non-Gaussian states.
These methods are applied to cosmology in Section~\ref{s:TwoField}, focusing
on Higgs-like inflation. The results are, however, more general and can easily
be adapted to any potential. This section will demonstrate the importance of
going beyond Gaussian dynamics, including higher-order moments, and
maintaining non-adiabatic regimes.  A detailed cosmological analysis,
including numerical simulations and analytical approximations, is performed in
Section~\ref{test}, where observational implications are discussed. The
derivations in the present paper justify the more concise physical discussion
presented in \cite{Inflation}.

\section{Canonical effective potentials}
\label{s:EffPot}

Our construction is based on canonical effective methods for non-adiabatic
quantum dynamics, which in a leading-order treatment has appeared several
times independently in various fields
\cite{VariationalEffAc,GaussianDyn,EnvQuantumChaos,QHDTunneling,CQC,CQCFieldsHom},
including quantum chaos, quantum chemistry, and quantum cosmology, but has
only recently been worked out to higher orders using systematic methods of
Poisson manifolds \cite{Bosonize,EffPotRealize}. While higher orders go beyond
Gaussian dynamics, the leading-order effects are closely related to Gaussian
approximations and can therefore be used for an illustration of the method.

\subsection{Relation to the time-dependent variational principle}

In order to illustrate our claim that quantum fluctuations can provide an
independent degree of freedom that can influence the inflationary dynamics, we
first consider a canonical formulation of the time-dependent variational
principle for Gaussian states.

The most general parametrization of Gaussian fluctuations around the
homogeneous field $\psi$ can be represented by the wave function
\cite{VariationalEffAc}
\begin{eqnarray} 
\Psi(\psi' | \psi, \pi_{\psi},\varphi, \pi_{\varphi})&=&\frac{1}{\left(2
    \pi \varphi^2\right)^{1/4}}
\exp\left(-{\textstyle\frac{1}{4}}\varphi^{-2}(1- 2
i \varphi  \pi_{\varphi})(\psi'-\psi)^2\right)\nonumber\\
&&\qquad\qquad\times \exp(i \pi_{\psi}
  (\psi'-\psi))\exp(-{\textstyle\frac{1}{2}}i \varphi \pi_{\varphi}
  )\,. \label{Psi} 
\end{eqnarray}
The notation is such that $\Psi$ is a wave function depending on $\psi'$ for
any choice of the parameters $\psi$, $\pi_{\psi}$, $\varphi$ and
$\pi_{\varphi}$.  Despite its lengthy form, this variational wave function has
some useful properties: It is normalized, $\langle \Psi| \Psi\rangle =1$, and
has basic expectation values
\begin{equation}
\langle \Psi| \hat{\psi}  | \Psi\rangle
=\psi\quad,\quad
 \langle \Psi| \hat{\pi}_{\psi}  | \Psi\rangle
=\pi_{\psi} 
\end{equation}
and variances
\begin{equation}
\langle \Psi| (\hat{\psi}-\psi)^2 | \Psi \rangle
=\varphi^2\quad,\quad 
\langle \Psi | (\hat{\pi}_{\psi}-\pi_{\psi})^2  | \Psi \rangle
=\pi_{\varphi}^2+\frac{1}{4 \varphi^2}
\end{equation}
where operators are defined with respect to the dependence of $\Psi$ on
$\psi'$. Moreover, $\Psi$ obeys  the conditions
\begin{eqnarray}
i \langle \Psi |\partial/\partial \psi| \Psi \rangle =\pi_{\psi}
\quad&,&\quad
i \langle \Psi |\partial/\partial \varphi| \Psi \rangle
=\pi_{\varphi} \\
\langle \Psi |\partial/\partial \pi_{\psi}| \Psi \rangle
=0\quad&,&\quad 
\langle \Psi |\partial/\partial \pi_{\varphi}| \Psi \rangle =0\,.
\end{eqnarray}

The equations of motion for the variational parameters, $\psi$, $\varphi$,
$\pi_{\psi}$ and $\pi_{\varphi}$, are given by the variation of the action
\begin{eqnarray}
S&=&\int {\rm d}t \left \langle \Psi
  \left|\left(i \partial_t-\hat{H}\right)\right| \Psi \right
\rangle\nonumber\\  
&=&\int {\rm d}t \left(i \dot{\psi} \left \langle \Psi
    |\partial/\partial \psi| \Psi \right \rangle+i
  \dot{\varphi} \left \langle \Psi |\partial/\partial
        \varphi| \Psi \right \rangle-\langle \Psi|\hat{H}|\Psi\rangle
\right) 
\end{eqnarray}
using the chain rule. The identities obeyed by $\Psi$ therefore allow
us to write the action in canonical form,
\begin{equation}
  S=\int {\rm d}t \left(\dot{\psi}\pi_{\psi}+\dot{\varphi}\pi_{\varphi}-H_{\rm
      G}\right) 
\end{equation}
where we defined the Gaussian Hamiltonian $H_{\rm G}=\langle\Psi|
\hat{H}|\Psi\rangle$. The variation of this action gives Hamilton's equations
\begin{equation}
\dot{\psi}=\frac{\partial H_{\rm G}}{\partial \pi_{\psi}}\quad,\quad
\dot{\pi}_{\psi}=-\frac{\partial H_{\rm G}}{\partial \psi}\quad,\quad
\dot{\varphi}=\frac{\partial H_{\rm G}}{\partial \pi_{\varphi}}\quad,\quad
\dot{\pi}_{\varphi}=-\frac{\partial H_{\rm G}}{\partial \varphi}\,.
\end{equation}

For example, if we consider the Hamilton operator
\begin{equation}
\hat{H}=\frac{1}{2}\hat{\pi}_{\psi}^2+M^4\left(1-2
  \frac{\hat{\psi}^2}{v^2}+\frac{\hat{\psi}^4}{v^4}\right)
\end{equation}
with the Higgs-like potential, the Gaussian Hamiltonian is
\begin{equation}\label{Hg}
H_{\rm G}=\frac{1}{2}\pi_{\psi}^2+\frac{1}{2}\pi_{\varphi}^2+\frac{1}{8
  \varphi^2}+M^4\left(1-2 \frac{\psi^2}{v^2}+\frac{\psi^4}{v^4}+6
  \frac{\psi^2 \varphi^2}{v^4}-2 \frac{\varphi^2}{v^2}+3
  \frac{\varphi^4}{v^4}\right) \,.
\end{equation}

\subsection{Canonical effective methods}

While the Gaussian approximation is useful in a wide range of applications a
more general class of states is relevant for our application to inflation
where non-Gaussianities should be included in the analysis. Canonical
effective methods \cite{EffAc,Karpacz} provide a good alternative because they
allow for generally non-Gaussian states while still retaining the canonical
structure that makes Gaussian states attractive. Importantly, it is not
required to find a specific representation of non-Gaussian states as wave
functions, which would be much more involved than (\ref{Psi}).  Instead, one
can formulate states of a quantum system in terms of expectation values and
moments assigned by a generic state to the basic operators $\hat{\psi}$ and
$\hat{\pi}_{\psi}$. The evolution of a state is then formulated as a dynamical
system for the basic expectation values $\psi=\langle\hat{\psi}\rangle$ and
$\pi_{\psi}=\langle\hat{\pi}_{\psi}\rangle$ as well as the moments
\begin{equation}
 \Delta(\psi^a \pi_{\psi}^b)=\left \langle (\hat{\psi}-\langle
     \hat{\psi}\rangle)^a(\hat{\pi}_{\psi}-\langle
     \hat{\pi}_{\psi}\rangle)^b\right \rangle_{\rm Weyl} \,,
\end{equation}
using Weyl (or completely symmetric) ordering in order to avoid overcounting
degrees of freedom.

The basic expectation values and moments inherit a Poisson structure from the
commutator,
\begin{equation}\label{Poisson}
\left\{\langle \hat{A}\rangle,\langle \hat{B}\rangle\right\}=\frac{1}{i
  \hbar}\left\langle [\hat{A},\hat{B}]\right\rangle\,,
\end{equation}
augmented by the Leibniz rule in an application to moments. The equations of
motion for some phase space function, $F(\psi,\pi_{\psi},\Delta(\cdot))$, are
then given in the form of the usual Hamilton's equations,
\begin{equation}
 \dot{F}(\psi,\pi_{\psi},\Delta(\cdot))=\left\{F,H_{\rm Q}\right\}
\end{equation}
with a quantum Hamiltonian $H_{\rm Q}=\langle\hat{H}\rangle$ defined as the
expectation value of the Hamilton operator $\hat{H}$ in a generic (not
necessarily Gaussian) state.  For a Hamiltonian of the form
$\hat{H}=\frac{1}{2}\hat{\pi}_{\psi}^2+\hat{V}(\psi)$, this definition implies
the quantum Hamiltonian
\begin{equation}
H_{\rm Q}=\langle \hat{H}\rangle=
\frac{1}{2}\pi_{\psi}^2+\frac{1}{2}\Delta(\pi_{\psi}^2)+
V(\psi)+\sum_{n=2}^{\infty}\frac{1}{n!}\frac{\partial^{n} 
  V}{\partial \psi^n}\Delta(\psi^n)\,. 
\end{equation}

The formulation of the system in terms of expectation values and moments
allows for a systematic canonical analysis at the semiclassical level. Written
directly for moments as coordinates on the quantum phase space, the Poisson
structure, based on (\ref{Poisson}) together with the Leibniz rule, is rather
complicated.  For instance, one can see that the Poisson bracket of two
moments is not constant and not linear in general
\cite{EffAc,HigherMoments}. Using moments as coordinates on a phase space
therefore leads to a more complicated inflationary analysis lacking a clear
separation between configuration and momentum variables. It is then unclear
how to determine kinetic and potential energies or a unique relationship
between specific phenomena and individual degrees of freedom.

In order to make the semiclassical analysis more clear, it is preferable to
choose a coordinate system that puts the Poisson bracket in canonical form as
in the variables used in \eqref{Hg}, but possibly extended to higher orders in
moments. The Darboux theorem \cite{Arnold} or its extension to Poisson
manifolds \cite{Weinstein} guarantees the existence of such coordinates, but
explicit constructions are in general difficult. For second-order moments, the
moment phase space is 3-dimensional and can be handled more easily than in the
general context. In this case, a canonical mapping has been found several
times independently
\cite{VariationalEffAc,GaussianDyn,EnvQuantumChaos,QHDTunneling}. It is
accomplished by the coordinate transformation
\begin{equation}\label{2mapping}
\Delta(\pi_{\psi}^2)=\pi_{\phi}^2+\frac{U}{\varphi^2}\quad,\quad
\Delta(\psi \pi_{\psi})=\varphi \pi_{\varphi}\quad,\quad
\Delta(\psi^2)=\varphi^2
\end{equation}
where $\left\{\varphi,\pi_{\varphi}\right\}=1$. The parameter
$U=\Delta(\psi^2)\Delta(\pi_{\psi}^2)-\Delta(\psi \pi_{\psi})^2$ is a
conserved quantity (or a Casimir variable of the algebra of second-order
moments), restricted by Heisenberg's uncertainty relation to obey the
inequality $U\geq \hbar^2/4$.  Direct calculations show that the
transformation (\ref{2mapping}) is a canonical realization of the algebra of
second-order moments. At this stage we already have a departure from the
Gaussian states, because the uncertainty for a pure Gaussian equals
$\hbar^2/4$, while we retain the uncertainty as a free (but bounded)
parameter.

Additional non-Gaussianity parameters, relevant for inflation, are revealed by
an extension of the canonical mapping to higher-order moments.  Considering
higher order semiclassical corrections implies more canonical degrees of
freedom. (For a single classical degree of freedom, the moments up to order
$N$ form a phase space of dimension $D = \sum_{j=2}^N(j+1)= \frac{1}{2}(N^2 +
3 N - 4)$.)  A canonical mapping for these higher-order semiclassical degrees
of freedom has only recently been derived in \cite{Bosonize,EffPotRealize} up
to the fourth order. For the relevant moments, the results are
\begin{eqnarray}\label{4omapping}
\Delta(\pi_{\psi}^2)&=&\sum_{i=1}^{5}\pi_{\varphi_i}^2+\sum_{i>j}
\frac{U}{(\varphi_i-\varphi_j)^2}\\ 
\Delta(\psi^2)&=&\sum_{i=1}^5\varphi_i^2\\
\Delta(\psi^3)&=&C \sum_{i=1}^5\varphi_i^3\\
\Delta(\psi^4)&=&C^2\sum_{i=1}^5\varphi_i^4+\sum \varphi_i^2 \varphi_j^2
\end{eqnarray}
while all other moments up to fourth order can be derived from the relevant
ones using suitable Poisson brackets. There are now five canonical pairs,
$(\varphi_i,\pi_{\varphi_i})$ and two Casimir variables, $U$ and $C$, forming
a 12-dimensional phase space of moments.

In order to parametrize the entire fourth-order semiclassical phase space we
had to introduce a total of five pairs of canonical degrees of freedom and two
Casimir variables, $U$ and $C$. In principle, we could consider all ten
non-constant semiclassical degrees of freedom, but in order to keep the
analysis simple, we take inspiration from some more terrestrial applications
\cite{QHD,EffPotRealize,Closure} and choose a moment closure, thereby
approximating higher-order moments in terms of lower-order ones. In
particular, we choose $\Delta(\pi_{\psi}^2)=\pi_{\varphi}^2+U/\varphi^2$,
$\Delta(\psi^2)=\varphi^2$, $\Delta(\psi^3)=a_3$ (or, alternatively, $a_{3}
\varphi^3$) and $\Delta(\psi^4)=a_4 \varphi^4$. This closure corresponds to
\eqref{4omapping} written in higher dimensional spherical coordinates with the
assumption that the angular momenta are small enough to be ignored.  The
parameter values $U=\hbar^2/4$, $a_3=0$ and $a_4=3$ correspond to the
Gaussian case. We can therefore think of this closure as describing the
non-Gaussianities by three parameters, $U$, $a_3$ and $\delta=a_4-3$, while
maintaining the same number of degrees of freedom as in the Gaussian case.

Considering a Higgs-inspired matter field coupled to a classical and isotropic
space-time background with spatial metric $h_{ij}=a(t)^2\delta_{ij}$ in terms
of proper time $t$, the standard Lagrangian
\begin{equation}
 L=\int{\rm d}^3x \sqrt{\det h} \left(\frac{1}{2} \dot{\psi}^2-\frac{1}{2}
   h^{ij}\partial_i\psi\partial_j\psi-V(\psi)\right)
\end{equation}
is first reduced to homogeneous form by assuming spatially constant $\psi$ and
integrating:
\begin{equation}
 L_{\rm hom}= \frac{1}{2} a(t)^3V_0 \dot{\psi}^2- a(t)^3 V_0 V(\psi)\,.
\end{equation}
The new parameter $V_0$, defined as the coordinate volume of the spatial
region in which inflation takes place, does not have physical implications but
merely ensures that the combination $a(t)^3V_0$ represents the spatial volume
in a coordinate-independent way. (The value of $a(t)^3V_0$ would be determined
by the maximum length scale on which approximate homogeneity may be assumed in
the early universe just before inflation \cite{MiniSup,Infrared}.) This
Lagrangian implies the scalar momentum 
\begin{equation}
 \pi_{\psi}= \frac{\partial L_{\rm hom}}{\partial \dot{\psi}}= a(t)^3V_0
 \dot{\psi}
\end{equation}
such that the Hamiltonian is given by
\begin{equation}
 H=\frac{1}{2
  a(t)^3V_0}\pi_{\psi}^2+a(t)^3V_0 V(\psi)\,.
\end{equation}

Quantizing the scalar field, using our explicit potential (\ref{Cl}), the
Hamilton operator is
\begin{equation}
 \hat{H}=\frac{1}{2
  a(t)^3V_0}\hat{\pi}_{\psi}^2+a(t)^3V_0 M^4
\left(1-\frac{\hat{\psi}^2}{v^2}\right)^2\,,
\end{equation}
keeping the background scale factor $a(t)$ classical.  The closure we choose
here implies the reduced version
\begin{eqnarray} \label{Hclosure} 
H^{\rm closure}_{\rm Q} &=&\frac{1}{2
    a(t)^3V_0}\pi_{\psi}^2+\frac{1}{2
    a(t)^3V_0}\pi_{\varphi}^2+\frac{U}{2 a(t)^3V_0 \varphi^2}\\
  &&+a(t)^3V_0 M^4\left(1+\left(\frac{6
        \varphi^2}{v^4}-\frac{2}{v^2}\right)\psi^2+\frac{\psi^4}{v^4}- 2\frac{
      \varphi^2}{v^2}+\frac{a_4 \varphi^4}{v^4}+4\frac{a_3 \varphi^3
      \psi}{v^4}\right) \nonumber
\end{eqnarray}
of the quantum Hamiltonian. While parameterizing some higher moments through a
moment closure is required for a tractable model, keeping at least one quantum
degree of freedom, $\varphi$, independent is crucial for a description of
non-adiabatic phases. In this way, our quantum Hamiltonian goes beyond
effective potentials of low-energy type, in particular the Coleman--Weinberg
potential \cite{ColemanWeinberg}. As shown in \cite{CW}, it is possible to
derive the Coleman--Weinberg potential from a field-theory version of
(\ref{Hclosure}) if one minimizes the Hamiltonian with respect to
$\varphi$. This step eliminates all independent quantum degrees of freedom
and, in the traditional treatment, is equivalent to a derivative expansion
performed in addition to the semiclassical expansion also applied here. The
derivative expansion eliminates non-adiabatic effects, which are retained here
by keeping $\varphi$ independent.

The effective Hamiltonian (\ref{Hclosure}) is very similar to the Gaussian
Hamiltonian \eqref{Hg}, which also retains an independent quantum variable,
but it is more general because of the presence of the new parameters $U$,
$a_3$ and $a_4$. As shall be shown later, the characteristics of our
inflationary phase depend crucially on these parameters. In particular for a
Gaussian state, inflation never ends, but if we consider small
non-Gaussianities parametrized by $U$, $a_3$ and $a_4$, we can obtain a
phenomenologically viable inflationary phase. Moreover, these parameters are
determined by the quantum state of the early universe, and so constraining
them with data would shed light on the character of the quantum state of the
early universe.

\section{Two-field model}
\label{s:TwoField}

After our transformation to canonical moment variables, we can uniquely
extract an effective potential from (\ref{Hclosure}),
\begin{eqnarray}\label{v-eff}
\frac{1}{M^4} V_{\rm eff}(\psi,\varphi)
&=&  1+\dfrac{U}{2 M^4 a^6V_0^2 \varphi^2}
+\left(6\frac{\varphi^2}{v^4}-\frac{2}{v^2}\right)\psi^2 +
\frac{\psi^4}{v^4}  
-2\frac{\varphi^2}{v^2}+\frac{4
  a_3\psi}{v^4}+a_4\frac{\varphi^4}{v^4} \nn \\ 
&\approx& 1+2\(\frac{\varphi^2-\varphi_c^2}{\varphi_c^2}\)\frac{\psi^2}{v^2}
+\frac{4 a_3\, \psi}{v^4} +
\frac{\psi^4}{v^4}-\frac{2}{3}\frac{\varphi^2}{\varphi_c^2} 
+a_{4}\frac{\varphi^4}{v^4}\,,
\end{eqnarray}
where $\varphi_c^2:= v^2/3$. 
By construction, the second field, $\varphi$, represents the quantum
fluctuation associated with the classical field $\psi$. As explained earlier,
the additional parameters, $U$, $a_3$ and $a_4$ describe a possibly
non-Gaussian quantum state of the background inflaton.

\subsection{Initial conditions and the trans-Planckian problem}

In the second line of the equation, we ignored the $U$-term $U/\(2M^4a^6V_0^2
\varphi^2\)$ in an approximation valid for sufficiently large scale factors
(or, rather, averaging volumes $a^3V_0$).  The origin of this term is purely
quantum and represents a potential barrier that enforces Heisenberg's
uncertainty relation for the fluctuation variable $\varphi$. This term can be
easily ignored after a few $e$-folds of inflation, but at early times its
presence necessitates $\varphi$ to start out at large values. The subsequent
non-adiabatic phase will be crucial for our model, and therefore this term
alleviates our need to fine-tune the initial condition for $\varphi$.

The main effect of this repulsive term in the potential is to push out
$\varphi$ to large values to begin with, after which we are always able to
neglect it throughout inflation. The initial $\varphi$ obtained in this way is
indeed consistent with requirements on inflation models. In particular, we can
easily obtain the initial condition $\varphi > \varphi_c$ of hybrid inflation
\cite{HybridInfl}: We expect the initial $\varphi$ to be large and can
therefore restrict the effective potential \eqref{v-eff} to the term quartic
in $\varphi$, together with the $U$-term relevant at early times. This
restricted potential has a local minimum at
\begin{equation} \label{phimin}
	\varphi=\sqrt[^6]{\frac{Uv^4}{4a^6V_0^2M^4a_4}} \,.
\end{equation}
We do not know much about the volume $a^3V_0$ of the initial spatial region
that was meant to expand in an inflationary way, but in order to avoid the
trans-Planckian problem \cite{TransPlanck,TransPlanck2,TransPlanck3}, we
should require that $a^3V_0>\ell_{\rm P}^3$. This lower bound implies the
upper bound 
\begin{equation}
	\varphi_{\rm ini}< \frac{1}{\ell_{\rm P}} \sqrt[^6]{\frac{Uv^4}{4
a_4 M^4}}
\end{equation}
for (\ref{phimin}). For parameters of the order $v\sim \mathcal{O}(M_{\rm
  P})$ and $M^4\ll M_{\rm P}^4$, as common in hybrid models and used
in our analysis to follow, the upper bound on $\varphi_{\rm ini}$ is much
greater than $\varphi_c$.

\subsection{Waterfall: Phase transitions}

Our effective potential \eqref{v-eff}, depending on the classical field $\psi$
and its fluctuation, $\varphi$ is of the hybrid-inflation type.  These models
typically produce a blue-shifted tilt when one starts with a large $\varphi$
and small $\psi$ \cite{HybridInfl}. Inflation in this scenario essentially
relies on the near-constant vacuum energy of $\psi$.  However, there is an
alternative scenario in the same model, the so-called waterfall regime
\cite{Waterfall,WaterfallNumerical}, realized at a later stage in our model in
which $\varphi$ has moved to and stays close to a minimum while $\psi$
gradually inches away from its vacuum value that has by then become an
unstable equilibrium position.

As we will show, initial conditions for the waterfall regime to take place are
generated in our extension of the model by a non-adiabatic phase in which
$\varphi$ is still large.  The subsequent waterfall regime then generates a
significant number of $e$-folds and leads to a red-shifted tilt for a wide
range of parameters.  For this scenario to take place, it is important that
our effective potential differs from the traditional hybrid one in that we
have an $a_4\varphi^4$ term as well as a $Z_2$-breaking term $a_3\psi$, which
is assumed to be small but not exactly zero.  The latter term relieves us of
the burden of supplying a non-zero initial value for $\psi$, which is required
to start the dynamics of the waterfall regime, as we shall demonstrate
later. Because both new terms depend on state parameters in our semiclassical
approximation, the resulting description of inflation is characterized by an
intimate link between observational features and properties of quantum states.

Another difference with the traditional hybrid model is that the hierarchy
between our set of parameters is more rigid, leaving less room for tuning and
ambiguity and making our results more robust. The traditional potential has
three parameters which can be adjusted independently, while in our case only
two (non-state) parameters are independent. This is so because we do not have a
generic two-field model but rather a single-field model accentuated by its
quantum fluctuation. As opposed to the traditional hybrid model
\cite{Waterfall}, we have two phase transitions characterized by non-adiabatic
behavior, and the majority of $e$-folds are created in between.

\begin{figure}
	\centering
	\includegraphics[width=0.7\textwidth]{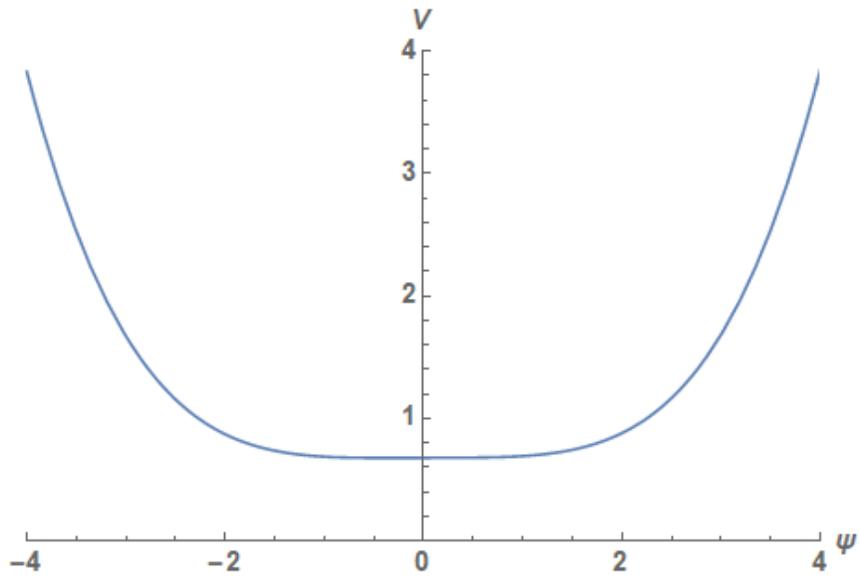}\vspace{1cm}
	\includegraphics[width=0.7\textwidth]{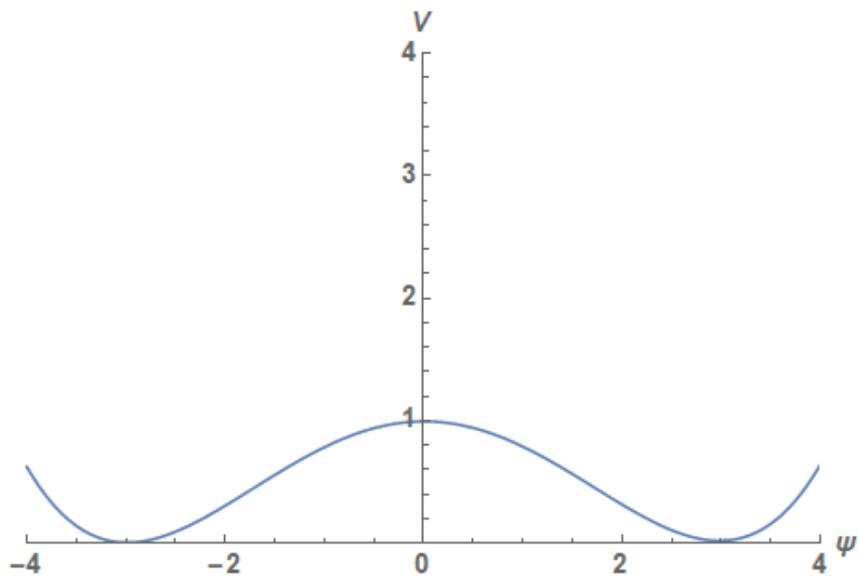}
	\caption{Shape of the potential $V(\psi)$ for constant $\varphi$ at
          early (top) and late times (bottom), defined relative to the time
          when $\varphi$ crosses $\varphi_c$.}
	\label{psi-trans}
\end{figure}
\begin{figure}
	\centering
	\includegraphics[width=0.7\textwidth]{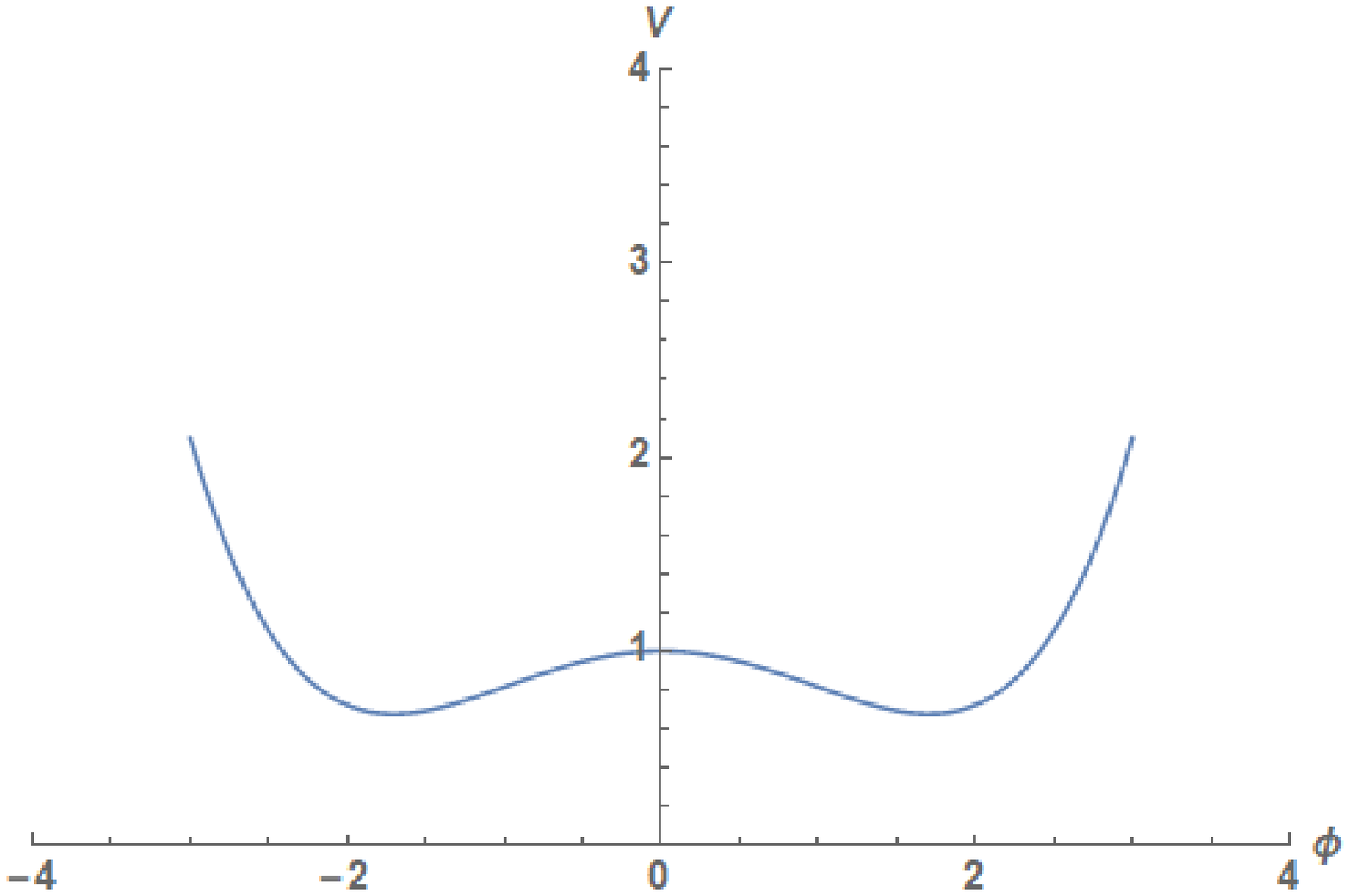}\vspace{1cm}
	\includegraphics[width=0.7\textwidth]{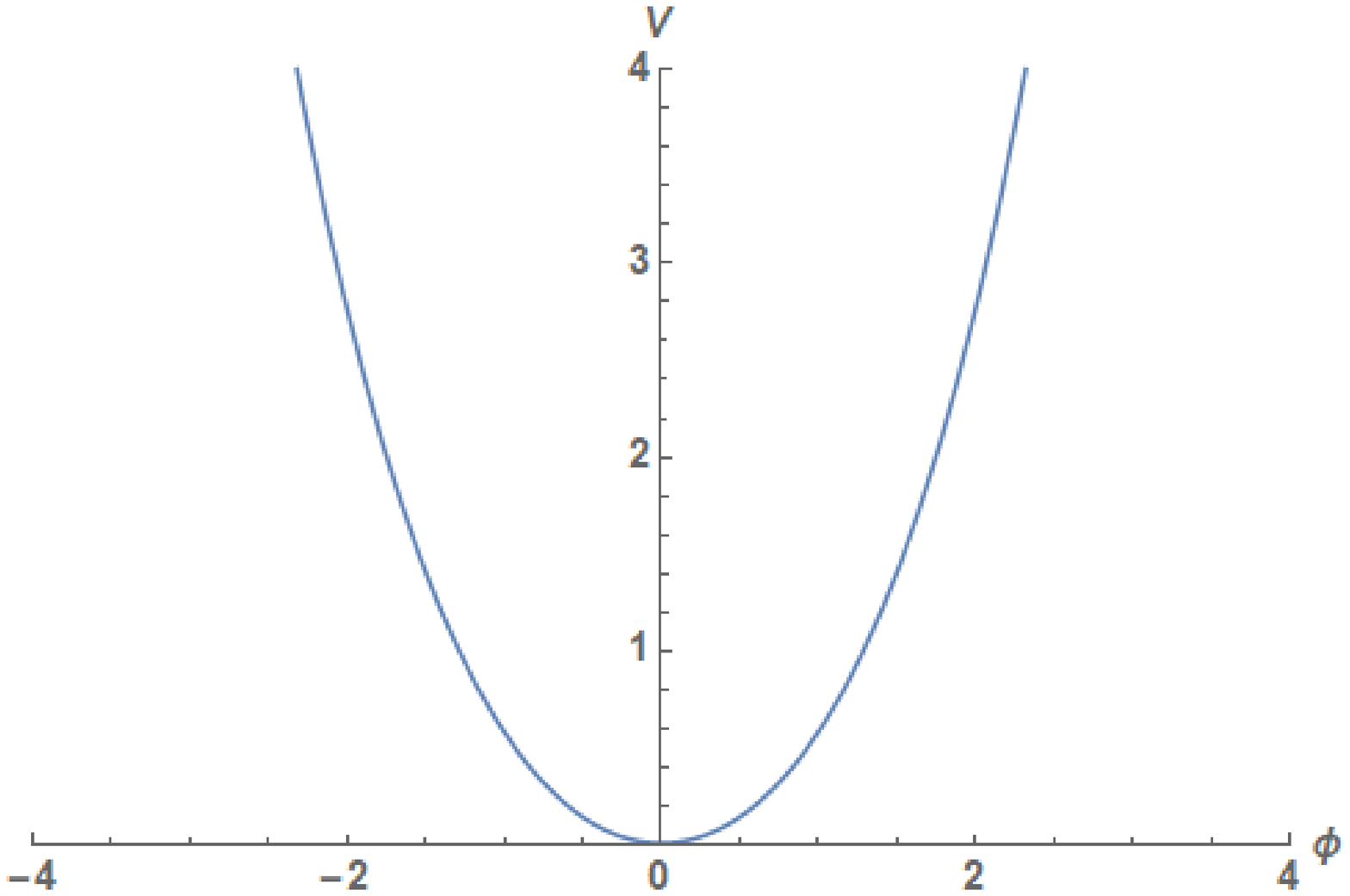}
	\caption{Shape of the potential $V(\varphi)$ for constant $\psi$ at
          early (top) and late times (bottom), defined relative to the time
          when $\varphi$ crosses $\varphi_c$.}
	\label{phi-trans}
\end{figure}

As in the original hybrid model, we start with some $\varphi>\varphi_c$ with
$\varphi$ quickly rolling down to its minima under an effective $\varphi^4$
term. This phase is driven by a simplified potential of the form
\begin{eqnarray}
\frac{V_{\rm eff}^{\varphi}}{M^4} = 1
-\frac{2}{3}\frac{\varphi^2}{\varphi_c^2} 
	+a_{4}\frac{\varphi^4}{v^4}
\end{eqnarray} 
since $\psi$ sits in its local minimum at the origin during this time and
therefore all $\psi$-terms can be ignored. Once $\varphi$ crosses $\varphi_c$,
the new true minima of $\psi$ are displaced from the origin due to a
tachyonic term in its effective potential, of the form
\begin{eqnarray}
\frac{V_{\rm eff}^{\psi}}{M^4} =
1+2\(\frac{\varphi^2-\varphi_c^2}{\varphi_c^2}\)\frac{\psi^2}{v^2} 
	+\frac{4 a_3 \psi}{v^4} +
        \frac{\psi^4}{v^4}-\frac{2}{3}\frac{\varphi^2}{\varphi_c^2} 
	+a_{4}\frac{\varphi^4}{v^4}\,,	
\end{eqnarray}

\begin{figure}
	\centering
	\includegraphics[width=0.75\textwidth]{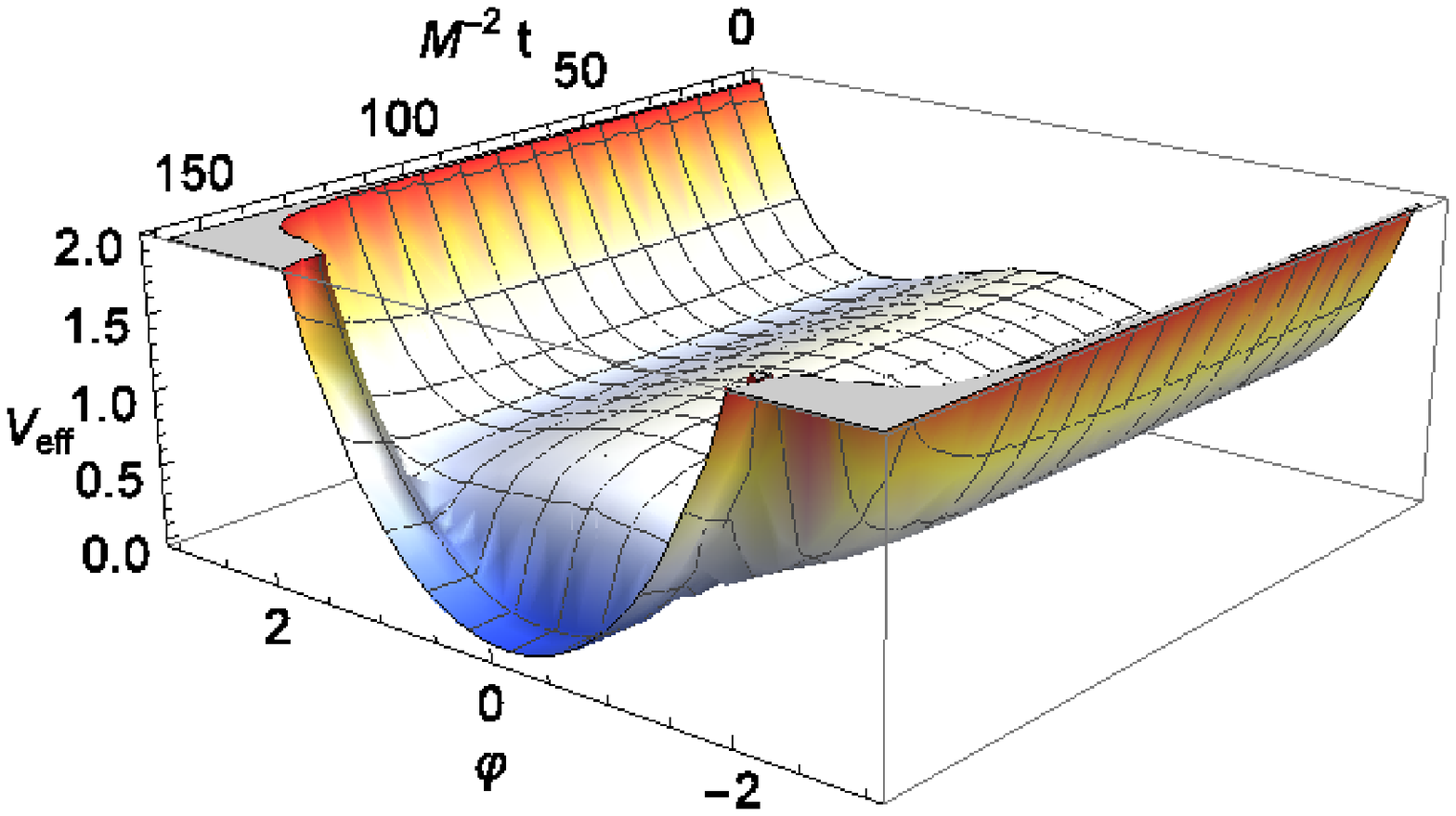}
	\caption{After a brief non-adiabatic phase when it rolls down a steep
          potential wall, $\varphi$ traces its
		minimum for the majority of inflation. The growth of $\psi^2$
		moves the $\varphi$-minima closer to zero, causing another
		non-adiabatic phase that ends with an approximate
		symmetry restoration for
                $\varphi$. The 
		parameters used are $v=3$, $a_3=0.05$ and $\delta=0.1$.
		\label{V-phi-3d}}
\end{figure}

\begin{figure}
	\centering
	\includegraphics[width=0.75\textwidth]{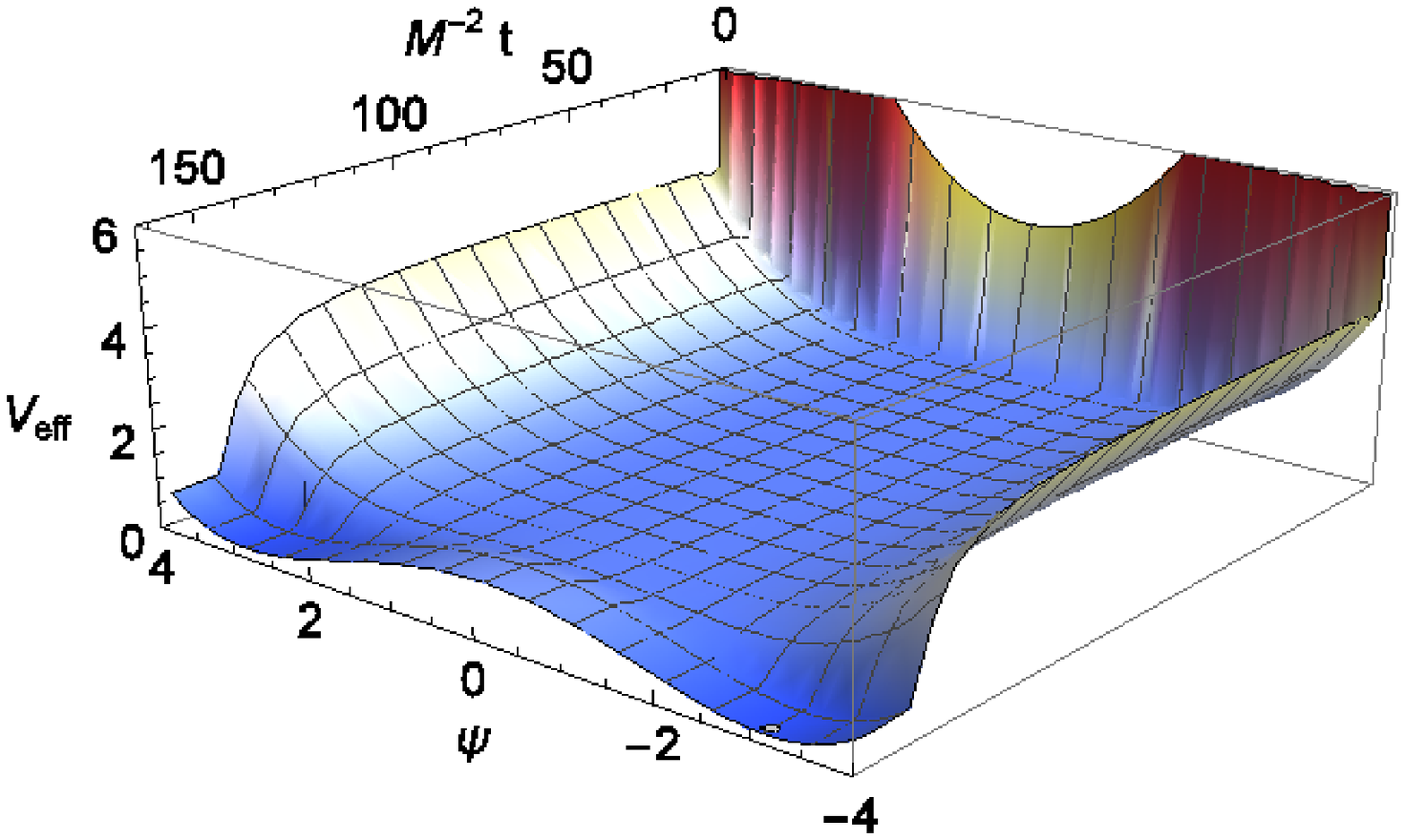}
	\caption{During the initial non-adiabatic phase, a phase transition
          akin to traditional 
		hybrid models occurs. Reflection symmetry in the potential
                is 
		slightly broken by the $a_3$-term (which is not
                apparent in the figure due to its smallness). 
		This non-Gaussianity term drives $\psi$ to its
                new stable point 
		where
		$\psi^2$ approaches $v^2$. The parameters are the same as
		Fig.~\ref{V-phi-3d}.
		\label{V-psi-3d}}
\end{figure}

Due to the $a_3$ term, the $Z_2$ symmetry of $\psi$ is broken and the field
starts slowly rolling away from the origin. This gradual change enables
$\varphi$ to closely follow its vacuum expectation value,
$\varphi_*$. Eventually, $\varphi_*$ approaches zero but never reaches
it due the uncertainty principle, 
thereby almost restoring the
symmetry for $\varphi$; this is the second phase transition mentioned
above. As shown in Figs.~\ref{psi-trans} and \ref{phi-trans}, $\varphi$ causes
the traditional phase transition when it crosses $\varphi_c$, and then the
slow roll of $\psi$ down its tachyonic hilltop will end in a second phase
transition. The whole process is clarified further by examining how the
effective potential changes in time, shown in Figs.~\ref{V-phi-3d} and
\ref{V-psi-3d}.

The hilltop phase generates the dominant number of $e$-folds, and it ends
automatically when $\psi$ reaches its new minimum. This is a new feature
compared to the traditional hybrid inflation and relies on the existence of a
$\varphi^4$ term in our effective potential. Our model is not a variant of the
original hybrid model \cite{Mutated_Hybrid}, such as the inverted-hybrid model \cite{Inv_Hybrid}
or a modified hilltop model \cite{Different_Hilltop}, or having corrections to 
the potential coming from supergravity-embedding of the model
\cite{SUGRA_Hybrid}; rather, we start with a Higgs-like model
and include effects from an initial quantum state that turn it into a hybrid
model with some additional terms.

\subsection{UV-completion and the swampland}

One of the conceptual requirements for inflation models is that they should
have a well-defined quantum completion. One way to implement this is to derive
specific forms of inflationary potentials from string theory constructions as
was done, for instance, in the case of natural inflation. Another recent
idea has been that of the swampland, a complement of the string landscape,
which stems from the fact that not all low-energy effective field theories can
be consistently completed in the ultraviolet into a quantum theory of gravity
\cite{Swampland,Swamp}. In order for an effective field theory to be
consistent, it would have to satisfy the eponymous swampland constraints. This
is a much more general way in which quantum gravity may restrict the form of
the potential, amongst other things, in the low-energy effective field theory
used as the starting point for inflation. More specifically, it has been
argued that many models of (at least) single-field inflation are not
consistent with the swampland conjectures since the latter require either a
large value for the slope of the potential, $|V'|/V > \mathcal{O}(1)$, or
large tachyonic directions, $V''/V < -\mathcal{O}(1)$ \cite{dSC}.

Taken together, these conjectures severely restrict the lifetime of metastable
(quasi-)de Sitter spacetimes that can be built from string theory. In order to
obtain an estimate for the numbers of order one that appear in one of them,
the so-called de-Sitter conjecture, one has to resort to fundamental
properties of quantum gravity such as the absence of eternal inflation
\cite{no_EI1,No_EI2} or the trans-Planckian censorship conjecture
\cite{TransPlanckSwamp,TransPlanckR}. The latter has put a more concrete bound
on the duration of inflation which, when combined with the observed power
spectrum, imposes severe constraints on the allowed models for inflation. It
has been shown that only hilltop type of models, which generically allow for a
small slow-roll parameter $\epsilon$ but a big $\eta$, are the ones that
survive amongst all single-field models unless one invokes additional degrees
of freedom as in non-Bunch Davies initial states or warm inflation. Even for
hilltop potentials, which seem to be the most compatible with the swampland,
one has to resort to an arbitrary steepening of the potential to end inflation
so as not to have too many $e$-folds since that would once again make the
model incompatible with the constraints. To date, there are no string theory
realizations of any such single-field potential that can abruptly stop
inflation after a finite amount of time.

The remarkable feature of our new model is that it is able to give a viable
inflationary cosmology as well as a graceful exit with a tachyonic
(p)reheating, all starting from a Higgs-like single-field potential as the
main input. We are using only standard quantum mechanics in a non-adiabatic
semiclassical approximation and do not have to rely on unknown features of
quantum gravity. In addition, by virtue of the fact that the classical field
$\psi$ plays the role of the inflaton relevant for observable scales, this
model is essentially of the hilltop type which has recently been shown to be
preferred by the swampland and to be able to ameliorate the $\eta$-problem
\cite{TCC_TWF}. Quantum effects imply that the single-field classical
potential is, upon quantization, no longer a single-field model that would
have to be tuned in order to avoid having too many $e$-folds of
inflation or require any additional mechanism to achieve stability against radiative corrections \cite{Kaloper_Hybrid}. Moreover, our detailed derivations below reveal that the model
maintains a large value of the slow-roll parameter $\eta$ throughout inflation
(in addition to a small $\epsilon$, as is usually the case for a prototype
hilltop model). Indeed, it is when the value of $\eta$ becomes too large that
inflation ends in this model, once again thanks to effects of quantum
fluctuations of the classical field (as opposed to a generic second
field). All of this is possible even though we start with a single-field model
with a monomial potential, but then take into account the effects of quantum
fluctuations in a systematical manner.

\section{Analysis} \label{test}

The effective Hamiltonian \eqref{Hclosure} describes a two-field model with
standard kinetic terms in an expanding universe and an interaction potential
similar to hybrid models. A numerical analysis can be applied directly to
Hamilton's equations for $\psi$ and $\phi$ generated by $H^{\rm closure}_{\rm
  Q}$, (\ref{Hclosure}), using suitable initial values. We will present such
solutions in comparison with a slow-roll approximation to be developed first.

\subsection{Slow-roll approximation}
For inflationary applications of \eqref{Hclosure}, we are interested in a long
phase of slow roll that can be generated by $\psi$ staying near its initially
stable and then metastable equilibrium position at $\psi=0$. As long as
$\psi^2\ll v^2$ and $\varphi^2\approx \varphi^2_*$ is near a local minimum, the
slow-roll approximation can be used and evaluated analytically. This phase is
adiabatic and therefore does not require all terms in \eqref{Hclosure} that
are implied by semiclassical methods for non-adiabatic quantum
dynamics. However, as we have already seen, the remaining terms are essential
in achieving suitable initial values for the slow-roll phase and to end it
early enough. Throughout this analysis, we will also assume small
background non-Gaussianity. As our results will show, this assumption is
justified by observational constraints on the spectral index.

Given these conditions, the slow-roll parameters can be approximated as
\begin{eqnarray}
	\epsilon_{\varphi}&\equiv& \frac{1}{2}M_{\rm P}^2\(\frac{V_{\varphi}}{V}\)^2
	\approx \frac{1}{2}M_{\rm P}^2\(\frac{M^4}{P}\)^2
	\(\frac{4\varphi}{3\varphi_c^2}\(1-\frac{3\psi^2}{v^2}\)-
\frac{4a_4\varphi^3}{v^4}\)^2\\ 
	\epsilon_{\psi}&\equiv& \frac{1}{2}M_{\rm P}^2\(\frac{V_{\psi}}{V}\)^2
	\approx \frac{1}{2}M_{\rm P}^2\(\frac{M^4}{P}\)^2
	\(\frac{4\psi}{v^2}
	\(\frac{\varphi^2}{\varphi_c^2}+\frac{\psi^2}{v^2}-1\)+
\frac{4a_3}{v^4}\)^2\\ 
\eta_{\varphi\varphi}&\equiv& M_{\rm P}^2\frac{V_{\varphi\varphi}}{V}=
-\frac{M^4}{P}
\(\frac{4}{3\varphi_c^2}\(1-\frac{3\psi^2}{v^2}\)-\frac{12a_4\varphi^2}{v^2}\)\\
\eta_{\psi\psi}&\equiv& M_{\rm P}^2\frac{V_{\psi\psi}}{V}=\frac{M^4}{P}
\frac{4}{v^2}\(\frac{\varphi^2-\varphi^2_c}{\varphi_c^2}+\frac{3\psi^2}{v^2}\)\\ 
\eta_{\psi\varphi}&\equiv& 
M_{\rm P}^2\frac{V_{\varphi\psi}}{V}=\frac{M^4}{P}\frac{8\psi\varphi}{v^2\varphi_c^2} \,,
\end{eqnarray}
where $V_{\varphi}=\partial V/\partial \varphi$ and $V_{\psi}=\partial
V/\partial \psi$, iterated for higher derivatives.  The constant $P$ is the
initial potential energy, evaluated when $\varphi\approx \varphi_c$ and
$\psi\approx 0$.  In the following we set $M_{\rm P}=1$.  We will see later
that small non-Gaussianity ensures that $\varphi^2/\varphi_c^2-1\ll 1$.  Along
with the adiabatic approximation for $\varphi$, this inequality can ensure
that $\epsilon_{\psi}$ and $\eta_{\psi\psi}$ are very small. However
$\eta_{\varphi\varphi}$ is not necessarily small, even though
$\ddot{\varphi}\ll 3H\dot{\varphi}$ and $\dot{\varphi}^2\ll V$.

Our equations of motion, under slow roll, then read
\begin{eqnarray}
	\label{phi-eom}
	\frac{3H\dot{\varphi}}{M^4}&=&
        \frac{4\varphi}{3\varphi_c^2}\left(1-\frac{3\psi^2}{v^2}\right) 
	-\frac{4a_4\varphi^3}{v^4}\\
	\label{psi-eom}
	\frac{3H\dot{\psi}}{M^4}&=& -\frac{4\psi}{v^2}
        \left(\frac{\varphi^2-\varphi_c^2}{\varphi_c^2}+
\frac{\psi^2}{v^2}\right)- 
\frac{4a_3}{v^4}\,. 
\end{eqnarray}
where we can make $M$ implicit by rescaling $t\rightarrow t/M^2$.
The regime covered by our approximations can be split into two phases followed
by an end phase.

\subsubsection{Phase 1}
 
In early stages, we have $\psi^2\ll v^2$ and can thus ignore the term
$3\psi^2/v^2$ in \eqref{phi-eom}. Therefore, the constant $\varphi^2\approx
\varphi^2_*\approx 3\varphi^2_c/a_4$ is a solution. Adiabaticity
ensures that we can expand the equation of motion around the critical point
$\varphi_*$ where $V_{\varphi}(\varphi_*)=0$:
\begin{equation}
	\dot{\varphi}\approx 
        -\frac{1}{3H}V_{\varphi\varphi}(\varphi_*)(\varphi-\varphi_*) \,. 
\end{equation}
Defining $\varphi':= {\rm d}\varphi/{\rm d}N$ where $N$ is the number of
$e$-folds, we obtain
\begin{equation}
	\varphi'\approx -\eta_{\varphi\varphi}(\varphi=\varphi_*,\psi\approx
        0) (\varphi-\varphi_*)\,. 
\end{equation}
For small non-Gaussianity, we have $a_4=3+\delta$ with $\delta\ll 1$.
Choosing the initial value $\varphi(0)=\varphi_c$ for Phase 1 therefore implies
\begin{equation} \label{phi1}
	\varphi_1(N)\approx
        \frac{\varphi_c\delta}{2a_4}\exp(-\eta_{\varphi\varphi}(\varphi_*,0)N)
+\varphi_*
        \,.  
\end{equation}
Note that small non-Gaussianity also implies
$\varphi_*^2=\varphi_c^2+O(\delta)+O(\psi^2)$.

We can expect $\varphi^2/\varphi_c^2-1\approx -\delta/a_4$ to be much bigger
than $\psi^2/v^2$ at early times.  This reduces the second equation of motion,
\eqref{psi-eom}, to
\begin{equation}
	\psi'\approx
        \frac{1}{P}\frac{4}{v^2}\left(\frac{\delta}{a_4}\psi-
\frac{a_3}{v^2}\right) 
\end{equation}
which is solved by
\begin{equation}
	\psi_1(N)\approx 
	-\frac{a_3a_4}{\delta v^2}
\left(\exp\left(\frac{4\delta}{v^2a_4P}N\right)-1\right)
\label{psi1-sol}
\end{equation}
for an initial $\psi_1$ at the origin.  To summarize, Phase 1 is characterized
mathematically by the possibility to ignore the $\psi^2/v^2$ terms in
\eqref{phi-eom} and \eqref{psi-eom}.

\subsubsection{Phase 2}

As $\psi$ moves away from its metastable position at $\psi=0$, the terms
$\psi^2/v^2$ in the equations of motion will eventually have noticeable
effects even while they may still be small. In particular, the local minimum
of $\varphi$ at
\begin{equation}
 \varphi_*(\psi(t))^2=\frac{v^4}{3\varphi_c^2a_4}
\left(1-\frac{3\psi(t)^2}{v^2}\right) 
\end{equation}
is then time-dependent. The solution for $\varphi$ in Phase 2 can therefore be
obtained directly from (\ref{phi1}) by inserting the time-dependent
$\psi$ and $\varphi_*$,
\begin{equation}
	\varphi_2(N)=\varphi_1(N)|_{\psi\rightarrow\psi(N)}\,,
\end{equation}
using the solution for $\psi(N)\equiv\psi_2(N)$ to be derived now.  As implied by
adiabaticity, we still have $\varphi^2\approx\varphi_*^2$, tracking the local
minimum.

Our phase now is described by the first two terms of \eqref{psi-eom} dominating
over the $a_3$-term.  Therefore,
\begin{eqnarray}
	\psi'&\approx& 
	-\frac{1}{P}\frac{4\psi}{v^2}
	\left(\frac{\varphi_*(\psi(t))^2-\varphi_c^2}{\varphi_c^2}+
\frac{\psi^2}{v^2}\right)      \nonumber  \\
	&=& \frac{1}{P}\frac{4\psi}{v^2}
	\left(\frac{\delta}{a_4}+\frac{2\psi^2}{v^2}+
	O(\delta\psi^2/v^2)\right)\,. \label{psip2} 
\end{eqnarray}
which is solved by
\begin{equation}
	\psi_2(N)\approx -\sgn(a_3)
	\sqrt{\frac{\delta}{(2a_4/v^2+\delta/ \psi_{\rm g}^{2})
\exp(-8\delta(N-N_{\rm g})/(v^2Pa_4))-2a_4/v^2}}        \,.
\label{psi2-sol}
\end{equation}
(Although $a_3$ does not appear in our approximate equation (\ref{psip2}), its
sign determines the direction in which $\psi$ starts moving as a consequence
of reflection symmetry breaking.)  Here, the subscript ``g'' denotes the value
of solutions at the ``gluing'' point of the two phases, defined as the point
where the cubic term in \eqref{psi-eom} is on the order of the $a_3$-term; see
Fig.~\ref{psi-eom-terms} below for an illustration.

\subsubsection{End phase}

Even though Phase 1 and Phase 2 are sufficient to describe the majority of
inflation, finding the point at which inflation ends requires a qualitatively
different approximation compared with the above two phases. The physics is
also quite different. To see this, note that if we extend the approximations
of Phase 2 too far, we arrive at two wrong conclusions. First, $\psi$ will
eventually cross the point $\psi^2=v^2/3$, such that the two minima of $V_{\rm
  eff}(\varphi)$ meet at $\varphi_*=0$. Second, this behavior causes $\varphi$
to approach zero, such that the field $\psi$ ends up at its new $V_{\rm
  eff}(\psi)$-minimum, $\psi_{\rm min}=-v$ (assuming $a_3$ is positive).  The
former ($\varphi\to0$) is forbidden by the uncertainty principle, embodied in
our $U$-term in $V_{\rm eff}$ neglected so far in the slow-roll analysis, and
the latter is erroneous since it implies that once everything has settled,
$H^2$, which is proportional to $V_{\rm eff}$ during slow roll, would seem to
approach a negative value $4a_3\psi/v^4<0$.

However, this last conclusion certainly cannot be correct because our
classical potential (\ref{Cl}), a complete square $V_{\rm
  cl}(\psi)=M^4(1-\psi^2/v^2)^2$, is positive semidefinite. Therefore, it is
quantized to a positive, self-adjoint operator $\hat{V}$ which cannot possibly
have a negative expectation value $V_{\rm eff}=\langle\hat{V}\rangle$ in any
admissible state. In terms of moments used in our canonical effective
description, after $\psi$ crosses the value $v^2/3$, the fluctuation variable
$\varphi$ shrinks. Therefore, according to our moment closure introduced after
equation (\ref{4omapping}), the variance $\Delta(\psi^2)=\varphi^2$ as well as
the fourth-order moment $\Delta(\psi^4)=a_4\varphi^4$ approach zero, while
$\Delta(\psi^3)=a^3$ has so far been assumed constant. This latter assumption
violates higher-order uncertainty relations for small $\varphi$. 

We will not require a precise form of such higher-order uncertainty relations,
or a specific decreasing behavior of $\Delta(\psi^3)$ because, referring to
positivity, we know that the magnitude of the $a_3$-term in the potential is
not allowed to be larger than the sum of the rest of the terms in $V_{\rm
  eff}$. (But see the next subsection for numerical examples with decreasing
$\Delta(\psi^3)$.) This observation places an implicit bound on
non-Gaussianity parameters when our potential energy decreases at the end of
inflation.  Taking this outcome into account, our effective potential
eventually becomes
\begin{equation}
	\frac{V_{\rm eff}}{M^4}\approx 
	\left(1-\frac{\psi^2}{v^2}\right)^2
	+\frac{2}{3}\frac{\varphi^2}{\varphi_c^2}
\left(\frac{3\psi^2}{v^2}-1\right) 
	+\frac{U}{2M^4 a^6 V_0^2 \varphi^2}\,, 
	\label{}
\end{equation}
where we have neglected the $\varphi^4$ and $a_3$ terms for small
fluctuations.  The corrected values $\varphi_*$ of the two $\varphi$-minima
are now
\begin{equation}
	\varphi_*\approx \pm \left( \frac{u}{K(\psi^2)}\varphi_c^2
        \right)^{1/4}
\end{equation}
where
\begin{equation}
	u=  \frac{U}{M^4 a^6 V_0^2}\quad\mbox{and}\quad
	K(\psi^2)= \frac{4}{3}\left(\frac{3\psi^2}{v^2}-1\right)\,.
\end{equation}
Since $u$ is extremely small after $60$ $e$-folds, we have $|\varphi_*|\ll 1$.
The symmetry restoration for $\varphi$ is therefore only an approximate
one. In addition, we neglected the $O(\delta\psi^2/v^2)$-term in
\eqref{psip2}, but kept $\delta/a_3$. These two terms become comparable around
$\psi^2=v^2/3$ for our chosen parameters.  However, as we will see later in a
comparison with numerical solutions, setting $\varphi=0$ and using
the $\psi(N)$ expression of Phase 2 during the end phase gives a sufficiently
accurate number of $e$-folds.

\subsection{Comparison of analytical and numerical solutions}

\begin{figure}
	\centering	
	\includegraphics[width=0.8\textwidth]{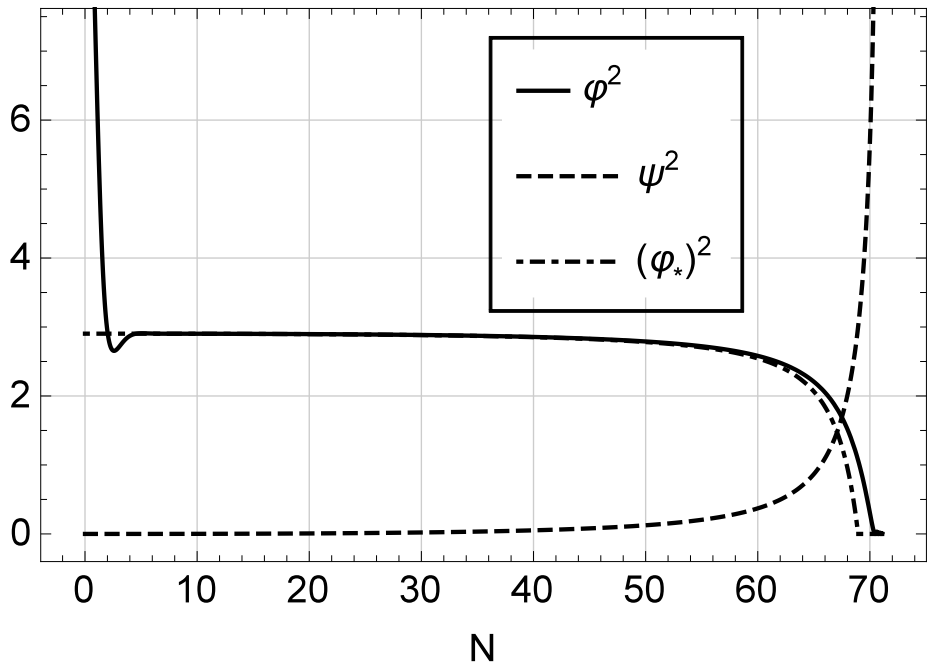}
	\caption{Overview of full numerical evolution. The field $\psi$
          remains small during inflation while $\varphi$ follows its vacuum
          expectation value $\varphi_*$ very closely throughout the whole
          evolution. After inflation ends, $\psi^2$ approaches $v^2$, a value
          cut off in this presentation. While the fields may take Planckian
          values, of the order one in natural units, except for very early
          times they hover near their potential minima where they imply
          sub-Planckian energy densities. Quantum-gravity effects are
          therefore negligible during inflation. The field $\psi^2$ increases
          at the end of inflation, but it merely approaches its new minimum
          seen in Fig.~\ref{V-psi-3d} and is not a run-away solution.}
	\label{general}
\end{figure}

\begin{figure}
	\centering	
	\includegraphics[width=0.7\textwidth]{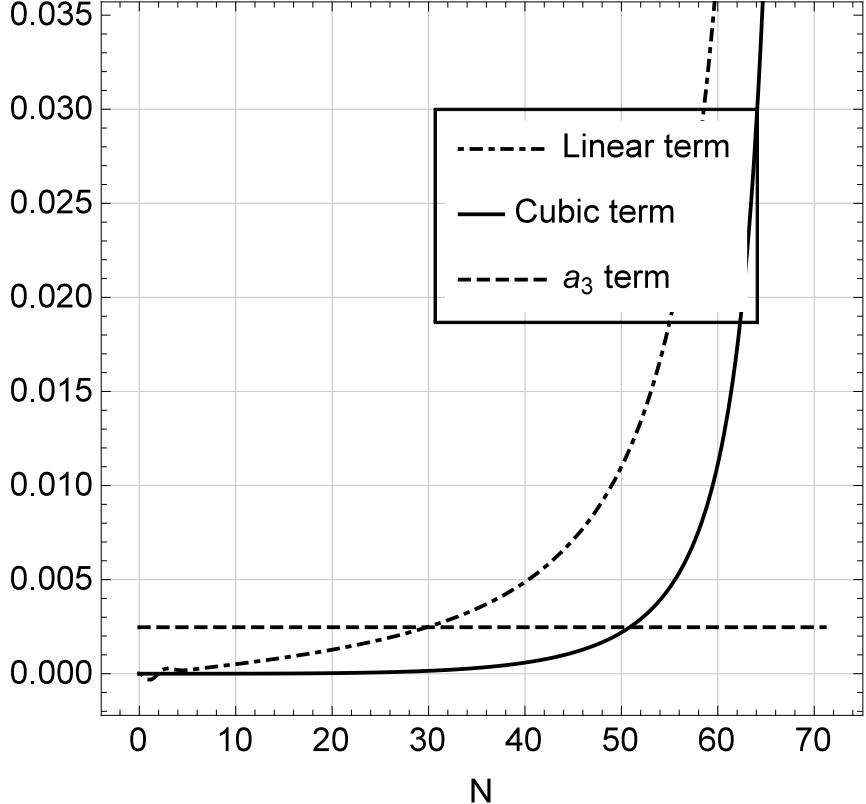}
	\caption{The magnitudes of individual terms in \eqref{psi-eom}
          as functions of $N$. The term $\psi^3/v^4$ in
          \eqref{psi-eom} approaches the order of $a_3/v^4$ around
          $N=50$, marking the transition point to Phase 2.}
	\label{psi-eom-terms}
\end{figure}

\begin{figure}
	\centering
            \includegraphics[width=0.7\textwidth]{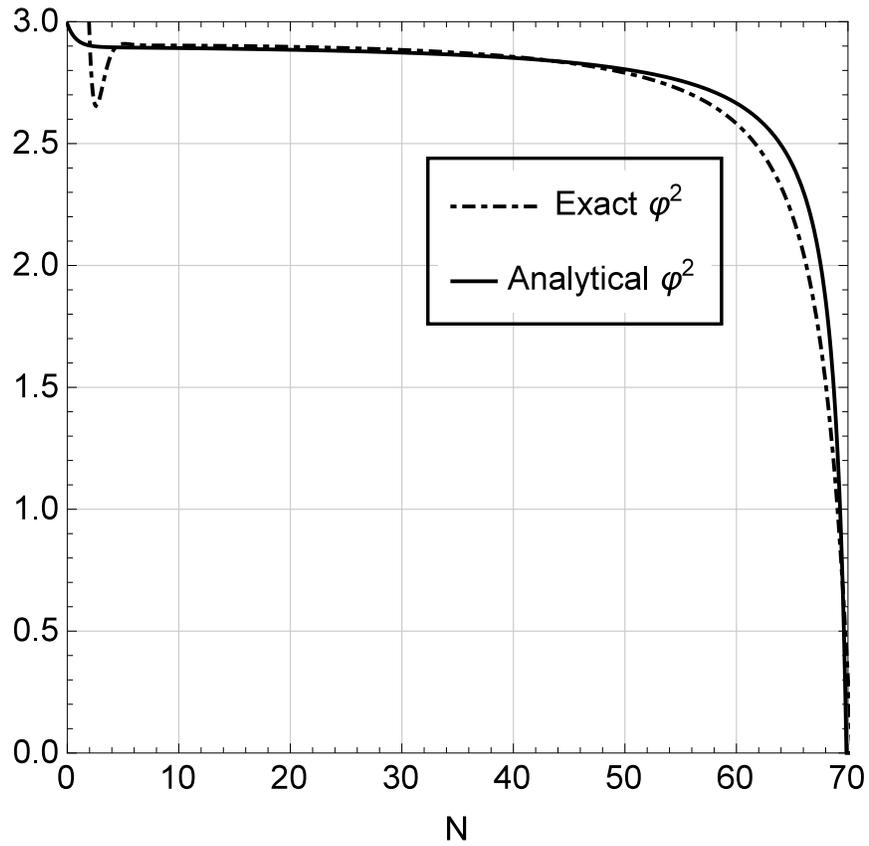}
            \caption{Comparison of analytical and numerical solutions for
              $\varphi(N)$. Our analytical solution for $\varphi(N)$
              agrees well with the full numerical one, justifying the
              adiabatic approximation during inflation. }
     \label{sol-comp1}
\end{figure}

\begin{figure}
	\centering
                   \includegraphics[width=0.7\textwidth]{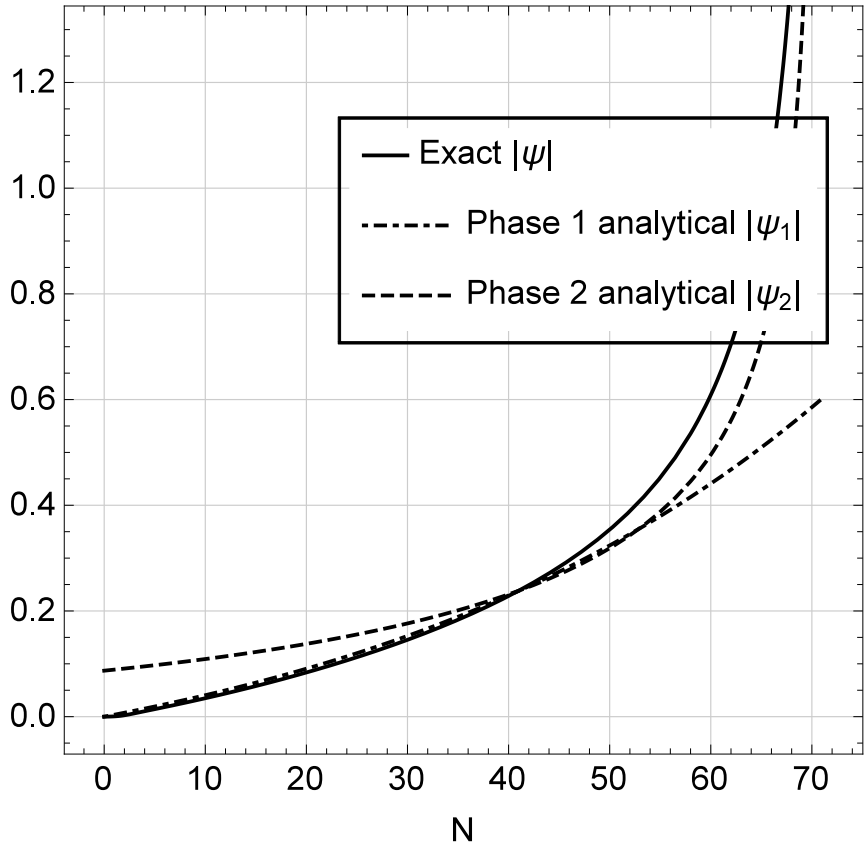}
            \caption{Comparison of analytical and numerical solutions for
              $\psi(N)$. The analytical solution agrees extremely
              well with the exact one in Phase 1 (before $N=50$), while small
              deviations occur in $\psi_2$ occur Phase 2 (after about
              $N=50$).}
     \label{sol-comp2}
\end{figure}

Our analytical solutions were obtained with certain approximations, but they
generally agree well with numerical solutions of the full equations,
\begin{eqnarray}
	\ddot{\varphi}+3H\dot{\varphi}&=& 
	\frac{4\varphi}{3\varphi_c^2}\left(1-\frac{3\psi^2}{v^2}\right)
	-\frac{4a_4\varphi^3}{v^4}\\
	\ddot{\psi}+3H\dot{\psi}&=& -\frac{4\psi}{v^2}
	\left(\frac{\varphi^2-\varphi_c^2}{\varphi_c^2}+
\frac{\psi^2}{v^2}\right)- 
\frac{4a_3}{v^4}\,, \label{psidd} 
\end{eqnarray} 
in situations relevant for inflation. To be specific, we choose parameters
$v=3$, $\delta=0.1$ and $a_3=0.05$ in our numerical
solutions. Figure~\ref{general} shows a representative example of full
numerical evolution. To test our analytical assumptions,
Fig.~\ref{psi-eom-terms} shows the magnitudes of individual terms that
contribute to the equation of motion (\ref{psidd}) for $\psi$, while
Figs.~\ref{sol-comp1} and \ref{sol-comp2} compare analytical and numerical
solutions of both equations.

\begin{figure}
	\centering	
	\includegraphics[width=0.7\textwidth]{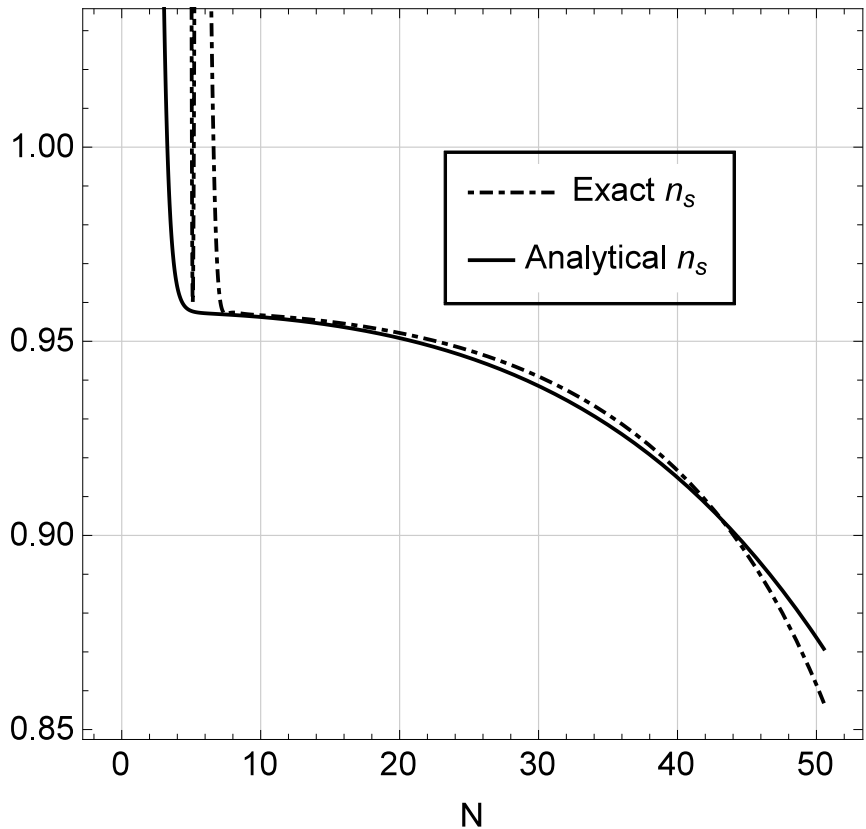}
	\caption{Analytical and numerical solutions for the spectral index
          $n_{\rm s}(N)$ in Phase 1.  Since Hubble exit takes place at least a
          $\Delta N\sim 60$ prior to the end of inflation, it can only occur
          in Phase 1. Importantly, $n_{\rm s}\approx 0.96$ at $\Delta N\sim 60$.}
	\label{ns-comp}
\end{figure}

\begin{figure}
	\centering	
	\includegraphics[width=0.7\textwidth]{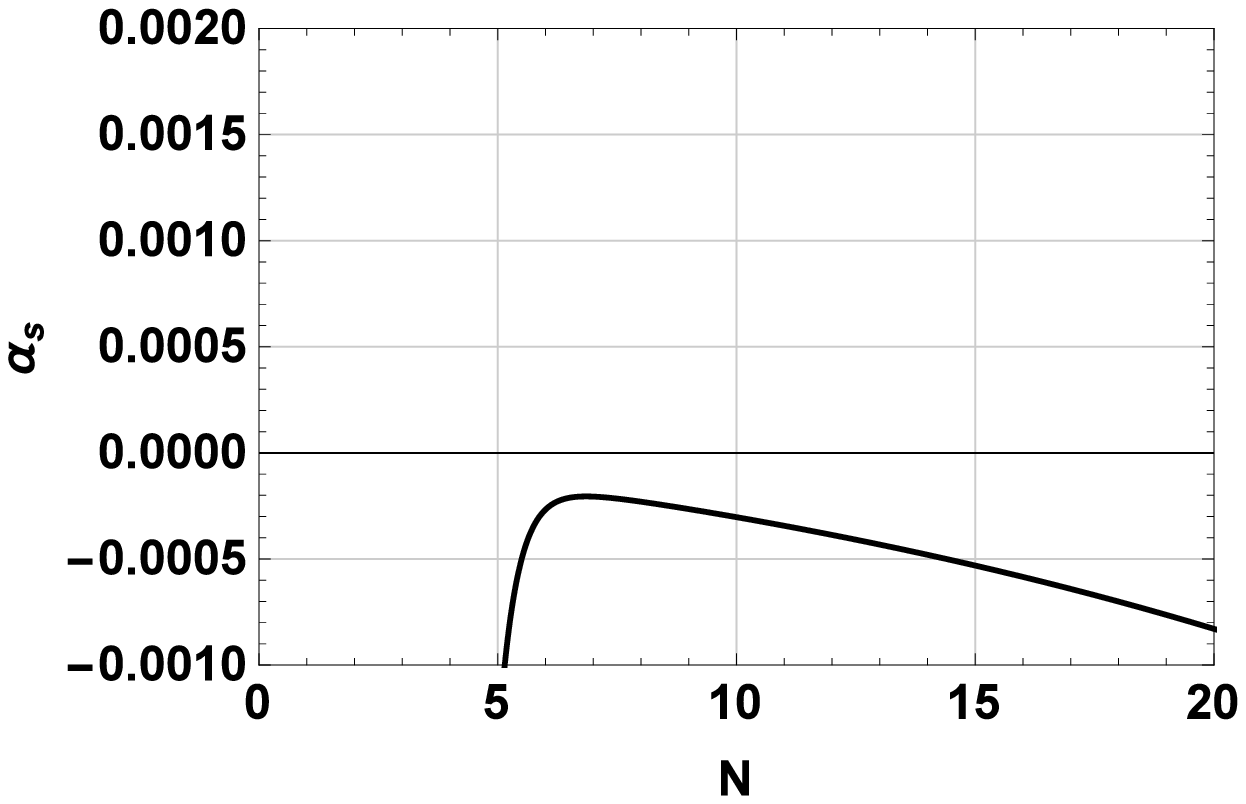}
	\caption{Analytical solution for the running $\alpha_{\rm s}\approx
          {\rm d}n_{\rm s}/{\rm d}N$ \cite{Lehners_2015} at early times, using
          a non-Gaussianity parameter $a_3=0.05$. Estimating Hubble exit
          at $N\sim 10$, $\alpha_{\rm s}$ is well within Planck's upper bound
          on the magnitude ($\sim 10^{-3}$).}
	\label{alpha-s-analytical}
\end{figure}

Cosmological parameters relevant for inflation are shown in the next figures,
Fig.~\ref{eta-psi} for the slow-roll parameter $\eta_{\psi\psi}$ which
eventually ends inflation, Fig.~\ref{ns-comp} for the spectral index according
to both analytical and numerical solutions, as well as its running in
Fig.~\ref{alpha-s-analytical}. As shown by these figures, the paramaters
easily imply solutions compatible with observational constraints. It is also
shown how $\eta_{\psi\psi}$ increases at an opportune time to end inflation
with just the right number of $e$-folds in order to avoid the trans-Planckian
problem.

\begin{figure}
	\centering	
	\includegraphics[width=0.7\textwidth]{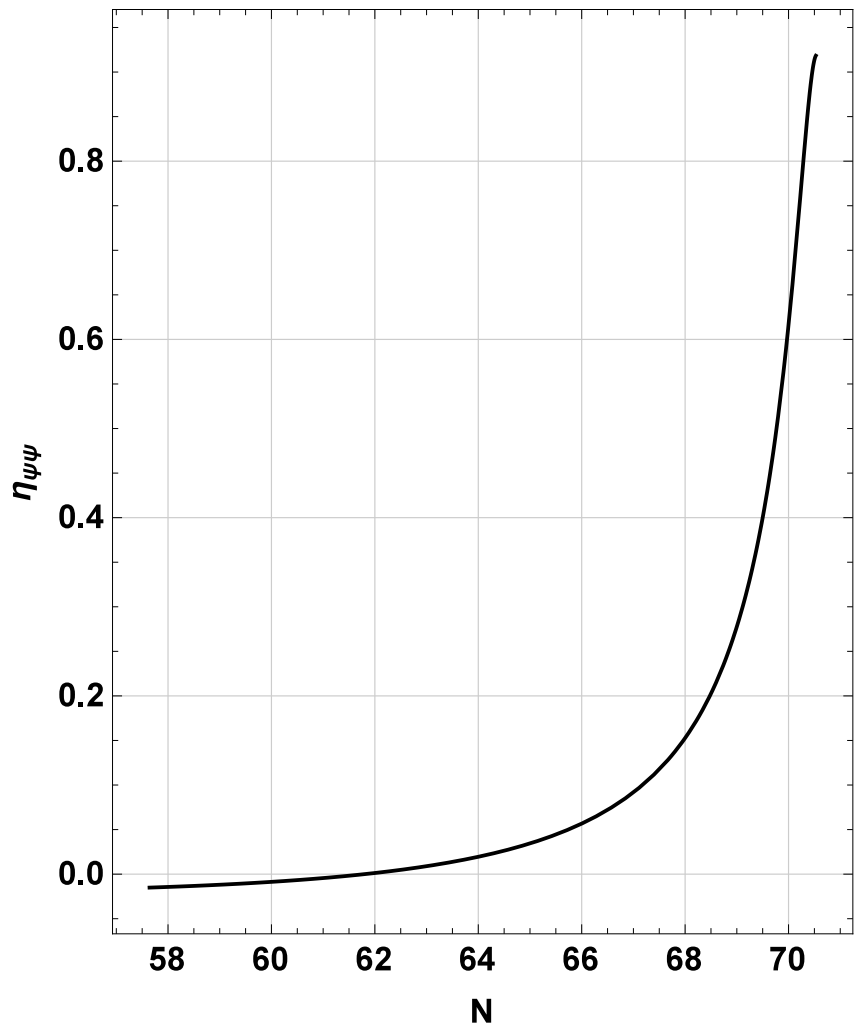}
	\caption{Late time behavior (Phase 2) of $\eta_{\psi\psi}(N)$ obtained
          from analytical solutions for $\psi(N)$ and $\varphi(N)$.  The
          slow-roll assumption starts being violated around $N\sim 70$,
          effectively ending inflation.}
	\label{eta-psi}
\end{figure}

\begin{figure}
	\centering	
	\includegraphics[width=0.9\textwidth]{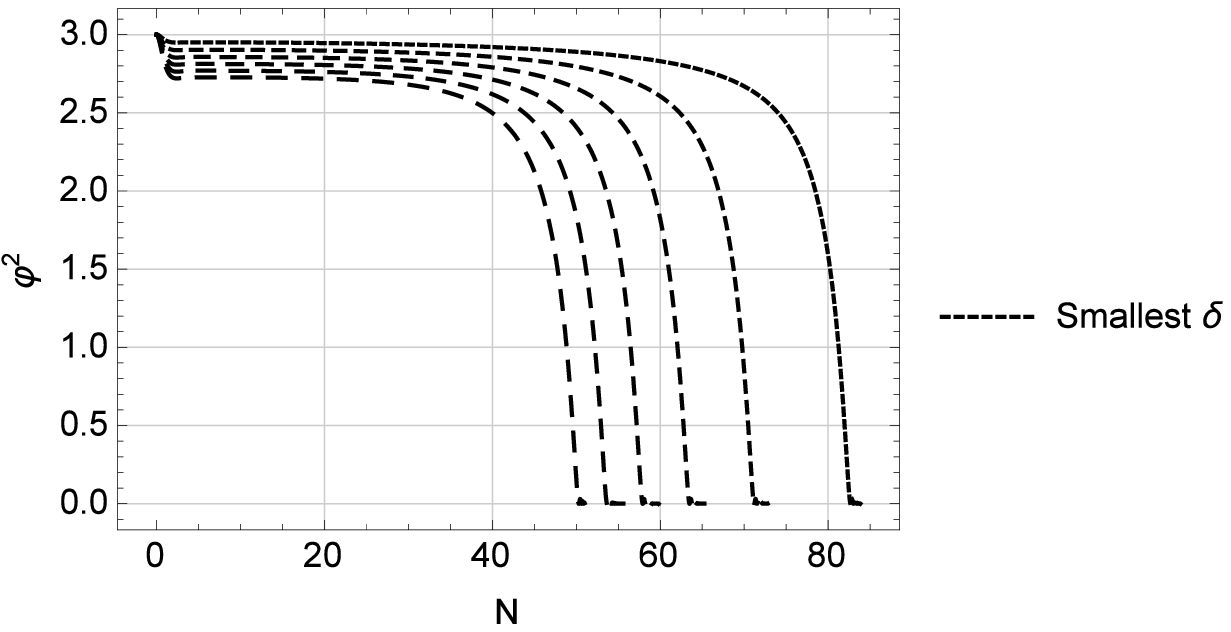}
	\caption{Evolution of $\varphi(N)^2$, from
          numerical solutions using $a_3=0.01\varphi^3$. Inflation ends
	  at $N_e$ where $\varphi(N_e)\approx 0$. Different curves
          correspond to different values of $a_4$, or $\delta=a_4-3$,
          where $\delta=0.05,0.1,0.15,0.2,0.25,3$. Smaller $\delta$ increase the
          duration of inflation.}
	\label{N-para}
\end{figure}

\begin{figure}
	\centering	
	\includegraphics[width=0.9\textwidth]{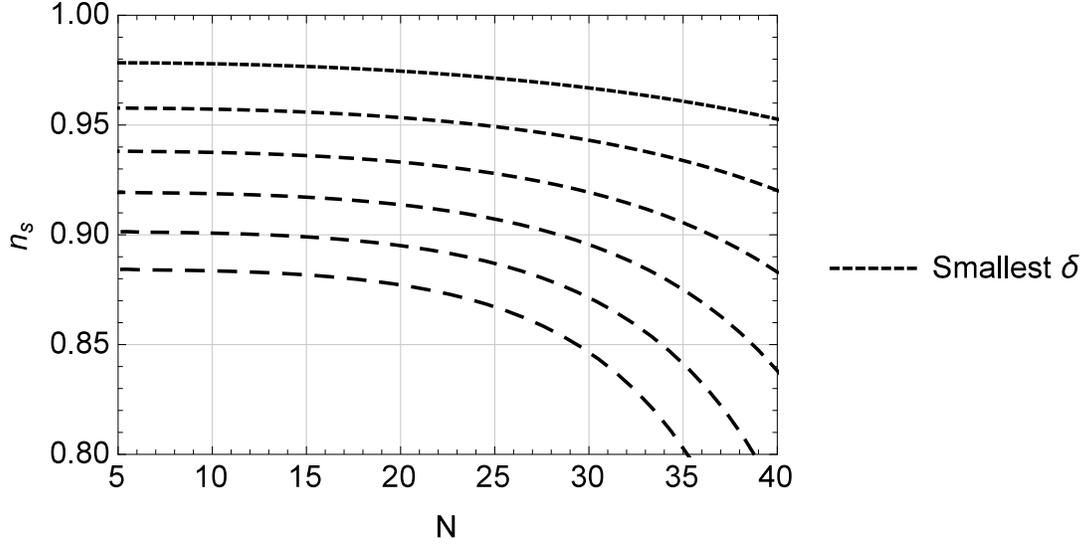}
	\caption{Spectral index $n_{\rm s}(N)$ as a function of $e$-folds $N$
	at Hubble exit
          from numerical solutions, using $a_3=0.01\varphi^3$. Different
          curves correspond to different values of $a_4$, or
          $\delta=a_4-3$, where
          $\delta=0.05,0.1,0.15,0.2,0.25,3$. Smaller $\delta$ brings the
          spectral index closer to one.}
	\label{ns-para}
\end{figure}

\begin{figure}
	\centering	
	\includegraphics[width=0.9\textwidth]{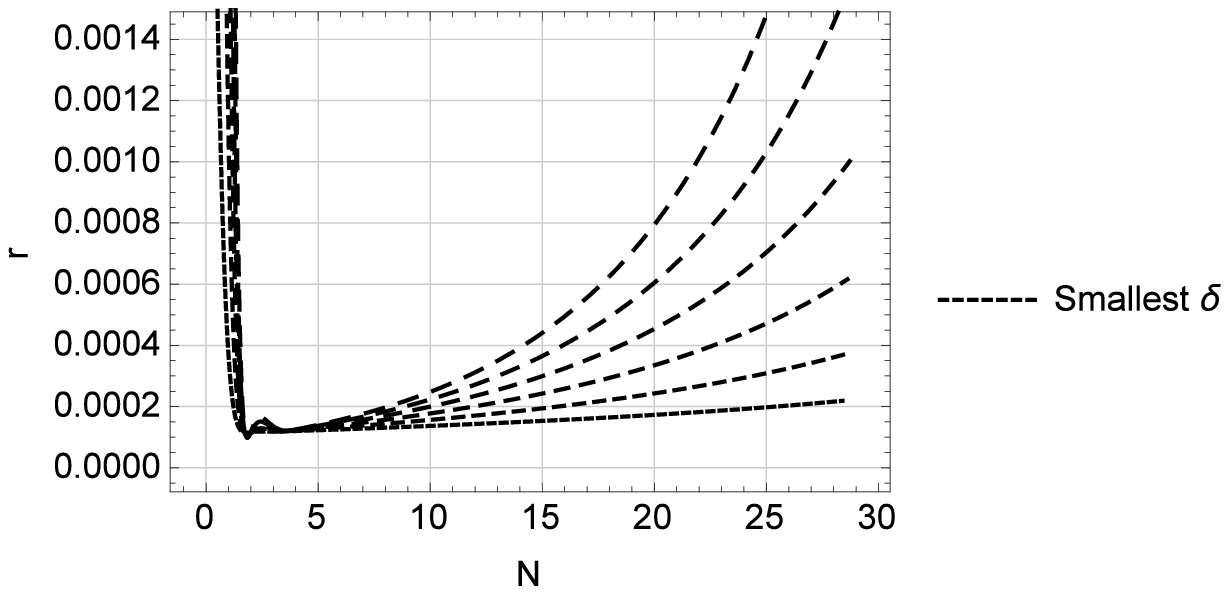}
	\caption{Tensor-to-scalar ratio $r(N)$ as a function of $e$-folds
		at Hubble exit
           from numerical solutions, using $a_3=0.01\varphi^3$. Different
          curves correspond to different values of $a_4$, or
          $\delta=a_4-3$, where
          $\delta=0.05,0.1,0.15,0.2,0.25,3$. Smaller $\delta$ decrease $r$.}
	\label{r-para}
\end{figure}

The role of non-Gaussianity parameters can also be studied. For instance,
parameterizing $a_3=0.01\varphi^3$ instead of a constant $a_3=0.05$ leads to
comparable results, as shown for the number of $e$-folds in
Fig.~\ref{N-para}. The effects of different choices of $\delta=a_4-3$
on the spectral index and the tensor-to-scalar ratio (computed as $r\approx
16\epsilon_{\sigma}$, $\sigma$ being the effective adiabatic field
\cite{Waterfall}) are shown in Figs.~\ref{ns-para} and \ref{r-para}.  An
important new result is that the non-Gaussianity parameters effectively
control the onset and duration of inflation, such that observationally
preferred numbers of $e$-folds can be obtained for reasonable choices of
background non-Gaussianity. In particular, only small deviations from a nearly
Gaussian ground state are required.

\subsection{Analytical results for cosmological observables}

Now we use the approximate analytical solutions to predict the number of
$e$-folds $N$ (starting from the crossing of $\phi=\phi_c$) and the spectral
index. In our model, both the classical field and its quantum fluctuation
undergo slow-roll evolution in different phases of the dynamics. Therefore,
they should both contribute to the curvature perturbation and one can write
down the effective adiabatic field $\sigma$ as a combination of both these
fields, $\psi$ and $\phi$.

In terms of the adiabatic field, consider the spectral index at around horizon
exit,
\begin{equation}
	n_s= 1-6\epsilon_{\sigma}+2\eta_{\sigma\sigma}\,.
\end{equation}
At early times, using the adiabatic approximation for
$\phi$ and small $\psi$, we have
\begin{equation}
	\epsilon_{\sigma}=\epsilon_{\psi}+\epsilon_{\varphi}
	\approx 0+ O(\psi^2,\delta^2,a_3^2)\,.
\end{equation}
For $\eta_{\sigma\sigma}$ we have \cite{Waterfall}
\begin{eqnarray}
	\eta_{\sigma\sigma}&=& \eta_{\varphi\varphi}\cos^2\theta
	+\eta_{\psi\psi}\sin^2\theta
	+2\eta_{\varphi\psi}\sin\theta\cos\theta
\end{eqnarray}
where $\theta$ is defined such that
\begin{equation}
 \cos\theta= \frac{\dot{\varphi}}{\sqrt{\dot{\varphi}^2+\dot{\psi}^2}}
	\quad,\quad
	\sin\theta= \frac{\dot{\psi}}{\sqrt{\dot{\varphi}^2+\dot{\psi}^2}}\,.
\end{equation}
Using the slow roll equations of motion for $\psi$ and $\dot{\varphi}\approx
\dot{\varphi}_*=-3\psi(a_4\varphi_*)^{-1}\dot{\psi}$ we obtain
\begin{equation}
	\cos\theta
	\approx -\frac{3\psi}{a_4\varphi_*}\sin\theta\quad,\quad
	\sin\theta
	\approx 1\,,
\end{equation}
where we used $V_{\psi}\gg V_{\varphi}\approx 0$.
To leading order of $\psi$, we therefore have
\begin{eqnarray}
	\eta_{\varphi\varphi}\cos^2\theta&\approx& 0+O(\delta^2,\alpha_3^2,\psi^2)\\
	\eta_{\varphi\psi}\sin\theta\cos\theta &\approx&0+O(\psi^2)\\
	\eta_{\psi\psi}\sin^2\theta &\approx&-\frac{4\delta}{a_4Pv^2}+O(\psi^2)\,,
\end{eqnarray}
%In phase 1, where horizon exit occurs, we have
%\begin{eqnarray}
%	\eta_{\psi\psi}&\approx& 
%		\frac{1}{V_0}(\frac{4}{v^2}
%		(-\frac{\delta}{a_4}-\frac{3\psi^2}{v^2})+\frac{12\psi^2}{v^4})\\
%		&=& -\frac{4\delta}{a_4V_0v^2}\,.
%\end{eqnarray}
such that
\begin{equation}
	n_{\rm s}\approx 1-\frac{8\delta}{a_4Pv^2}\,.
\end{equation}
Evaluating
\begin{equation}
	P\equiv V(\varphi_*(\psi=0),\psi=0)
  = 1-\frac{1}{a_4}
  \approx \frac{2}{3}+\frac{\delta}{9}+O(\delta^2)
\end{equation}
leads to the final expression
\begin{equation}
	n_{\rm s}\approx 1-12\frac{\delta}{a_4v^2}+O(\delta^2)\,.
	\label{ns}
\end{equation}
Imposing a slow-roll condition such as $\eta_{\psi\psi}\sim 10^{-2}$ requires
$v^2/\delta\sim O(10^2)$, which implies typical values of $n_{\rm s}$ in the range
$0.9<n_{\rm s}<1$.

Now, for total number of $e$-folds $N_e$ we first need to find the value
$\psi_e$ of $\psi$ at which inflation ends. Approximately, this stage occurs
when
\begin{equation}
	\eta_{\psi\psi}(\varphi_*,\psi_e)
	=\frac{V_{\psi\psi}}{V}|_{\varphi=\varphi_*,\psi=\psi_e}
	\approx 1 
\end{equation}
during the end phase. Under the approximation $\varphi\approx 0$, we have
\begin{equation}
	\frac{V_{\psi\psi}}{V}\approx
        \frac{4}{v^2}\frac{3\psi^2/v^2-1}{(1-\psi^2/v^2)^2}\,.  
\end{equation}
Then $V_{\psi\psi}\approx V(\varphi_*,\psi)$ gives
\begin{eqnarray}
	\frac{\psi_e^2}{v^2}&\approx& 1+\frac{6}{v^2}\pm
        2\sqrt{\frac{9}{v^4}+\frac{2}{v^2}}\\ 
	&=&1+\frac{6}{v^2}\left(1-\sqrt{1+\frac{2v^2}{9}}\right)\,,
\end{eqnarray}
where we chose the minus sign in the second line. From the above
expression we see that typically $\psi_e^2/v^2-1/3\sim O(10^{-1})$.  Then
using
\begin{equation}
	\Delta \psi \sim -\frac{V_{\psi}}{V}\Delta N \sim O(1) \Delta N\,,
\end{equation}
we see that that beyond $\psi^2/v^2=1/3$, we do not get many $e$-folds before
reaching the point $\eta_{\psi\psi}\approx 1$, effectively ending
inflation. In terms of the total number of $e$-folds, it is therefore
justified to approximate
\begin{equation}
	\psi_2(N)^2\approx v^2/3 \quad\mbox{such that}\quad \varphi_*^2=0
\end{equation}
as the end point of inflation.

\begin{figure}
    \centering
    \includegraphics[width=0.9\textwidth]{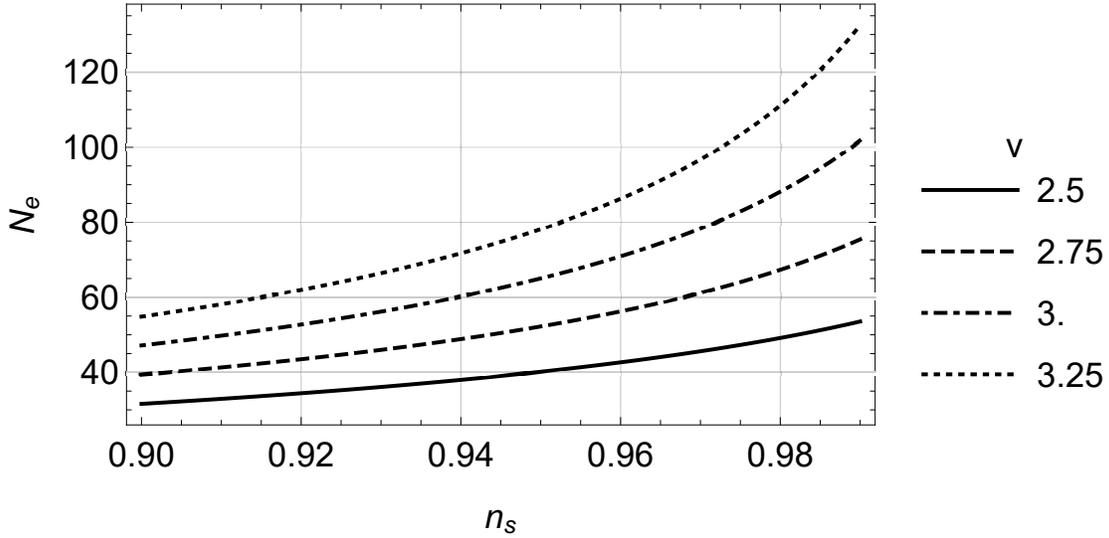}
    \caption{The number of $e$-folds, $N_e$, increases as a function of the
      spectral index $n_{\rm s}$, using the approximate relation
      \eqref{N-ns}. The function is shown for varying parameters $v$ in the
      potential, while $a_3=0.05$. As a function of the non-Gaussianity
      parameters, the number of $e$-folds decreases; see Fig.~\ref{N-alpha3}.
      (Note that in the 
      analytical relation (\ref{ns}), the variation of $n_{\rm s}$ mirrors
      the non-Gaussianity ratio
      $\delta/(a_4v^2)$.)
	    \label{N-ns-fig}}
\end{figure}

Since our analytical solution consists of $\psi_1$ and $\psi_2$, to find the
total number of $e$-folds $N_e$ at $\psi_2^2=v^2/3$, we must first find the
number of $e$-folds $N_{\rm g}$ at the gluing point. By definition of the latter,
\begin{equation}
	\psi_{\rm g}^3\equiv\psi_1(N_{\rm g})^3=-a_3 \,.
\end{equation}
Denoting 
\begin{equation}
	\eta\equiv |\eta_{\psi\psi}(\varphi_1,\psi_1)|\approx
        \frac{4\delta}{a_4Pv^2} 
	\approx \frac{6\delta}{a_4v^2}+O(\delta^2)\,,
\end{equation}
we have
\begin{equation}
2\eta=1-n_{\rm s}\,.
	\label{spectral-eta}
\end{equation}
Using \eqref{psi1-sol},
\begin{equation}
	\exp(\eta N_{\rm g})=\frac{v^2}{\psi_{\rm g}^2}\frac{\delta}{a_4}+1
\end{equation}
which, inserted in \eqref{psi2-sol}, using \eqref{spectral-eta}
and setting $\psi_2(N_e)^2=v^2/3$, implies
\begin{equation}
	N_e=\frac{1}{1-n_{\rm s}}\left(\log\left(\frac{2}{v^2}+\frac{1-n_{\rm
              s}}{12}\chi\right) 
	+2\log\left(\frac{1-n_{\rm s}}{12}\chi
        v^2+1\right)-\log\left(\frac{2}{v^2}+\frac{1-n_s}{4}\right)\right)\,, 
	\label{N-ns}
\end{equation}
where $\chi\equiv v^2/\psi_g^2$.
The relationship
(\ref{N-ns}) is illustrated in Fig.~\ref{N-ns-fig}.

\begin{figure}
    \centering
    \includegraphics[width=0.9\textwidth]{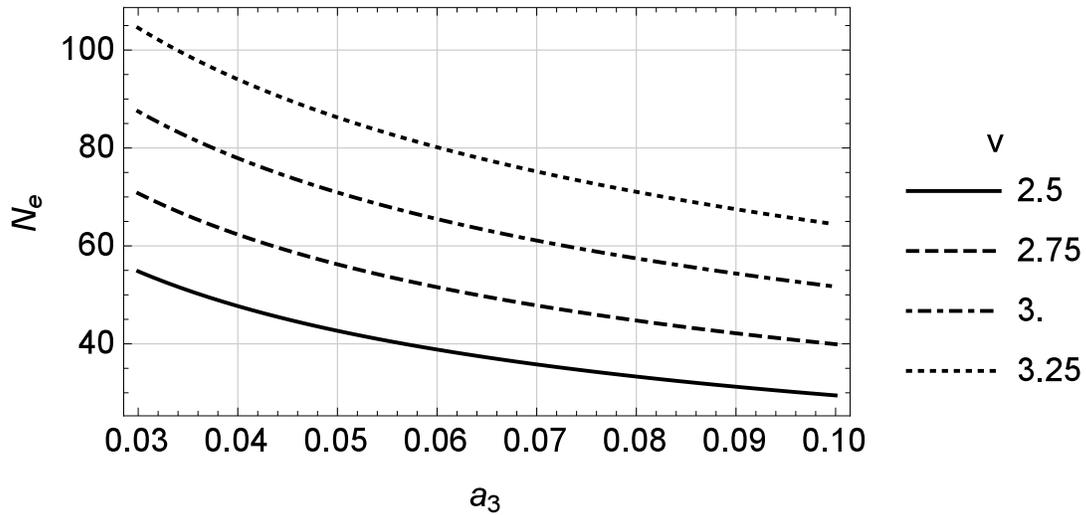}
    \caption{The number of $e$-folds, $N_e$, decreases with the amount of
      non-Gaussianity, 
	    parameterized by $a_3$, shown here for fixed $n_{\rm
              s}\approx 0.96$, $\delta=0.1$ and using \eqref{N-ns}. 
           Background non-Gaussianities increases the departure
	    from adiabatic evolution, effectively ending inflation earlier
            than  
	    desired.
	    \label{N-alpha3}}
\end{figure}

Aside from the parameter $v$ that appears in common Higgs-like or hybrid
models, our observables depend on two new parameters $a_3$ and $\delta$ which
describe the non-Gaussianity of the background state. Background
non-Gaussianity effectively controls the amount of non-adiabatic evolution due
to its modulation on the shifting of local $\varphi$-minima at $\varphi_*$.
The dependence of the number of $e$-folds on the non-Gaussianity parameter
$a_3$ is shown in Fig.~\ref{N-alpha3}, using the analytical solutions.

The dependence (\ref{N-ns}) of $N_e$ on $n_{\rm s}$ is more complicated than
in non-minimal Higgs models, but it is nevertheless related. To facilitate a
comparison, we rewrite the expression as
\begin{equation} \label{Nlog}
	N_e\approx
        \frac{f(1-n_{\rm s},v,a_3)}{1-n_{\rm s}}
\end{equation}
where the function $f$ describes a weak, logarithmic dependence on $1-n_{\rm
  s}$.  In non-minimal Higgs inflation, the analog of the function $f(1-n_{\rm
  s},v,\alpha_3)$ is constant ($f=2$) \cite{HiggsNonMin}. Here, the function
increases logarithmically with growing $1-n_{\rm s}$, taking values in the
range $1\lesssim f(1-n_{\rm s},v,a_3)\lesssim 5$ for typical parameter values
considered in our analysis. (An abbreviated derivation of (\ref{Nlog}) can be
found in \cite{Inflation}.)

\section{Conclusions}

Typically, potentials for the inflaton field are postulated so as to match
existing observations. On the other hand, one of the most remarkable successes
of inflation is that it explains the large-scale structure of the universe as
originating from quantum vacuum fluctuations. It is inconceivable to quantize
the fluctuations of the inflaton field \textit{alone} without taking into
account the quantum corrections to the background field potential. In other
words, one cannot simply express the inflationary potential in terms of
expectation values of the homogeneous background field, but should also take
fluctuations and higher moments of the quantum state into account. It is
customary to express the resulting effective potential in a derivative
expansion (of the Coleman-Weinberg type); however, this method is not
sufficient if one has to consider non-adiabatic evolution of the inflaton
field. Although a slow-rolling field does seem to justify an adiabatic
approximation, in this work we have shown how non-adiabaticity can play a
crucial role setting up the initial conditions for a slow roll phase as well
as help in ending it. We have presented a more general procedure for
calculating the effects of such non-adiabatic evolution in the context of
early-universe cosmology.

We have presented an observationally consistent extension of Higgs-like
inflation by introducing non-adiabatic quantum effects in a semiclassical
approximation although our formalism is applicable more generally for any
inflationary potential. As shown, these effects imply that the classical
potential is not only corrected in its coefficients but is also amended by new
terms for independent quantum degrees of freedom, in particular the quantum
fluctuation of the Higgs field. The original single-field model is therefore
turned into a multi-field model. The multi-field terms incorporate quantum
corrections of the background field, corresponding to backreaction of
radiative corrections. Since the single-field potential is renormalizable, our
quantum scenario is robust from the perspective of quantum field theory.

New interaction terms in the multi-field potential have coupling constants
that depend on the background state, parameterizing its non-Gaussianity. They
imply two new non-adiabatic phases that cannot be seen in low-energy
potentials or in cosmological studies based completely on slow-roll
approximations. In particular, an initial non-adiabatic phase, combined with
the uncertainty relation for the fluctuation degree of freedom, sets
successful initial conditions for inflation to take place, and a second
non-adiabatic phase ends inflation after the right number of
$e$-folds. Observational constraints show that background non-Gaussianity
should be small, but it must be non-zero for the non-adiabatic phases to be
realized. Our model is highly constrained because non-Gaussianity is bounded
from below, but we are nevertheless able to derive successful inflation in the
range of parameters available to us. 

Our model presents a new picture on the role of the quantum state in
inflationary cosmology. Quantum fluctuations do not only provide the seeds of
structure as initial conditions for perturbative inhomogeneity, they also play
a crucial role in guiding the inflationary dynamics of the background
state. With further analysis and observations, it may be possible to further
constrain the quantum state of the inflaton based on cosmological
investigations.

\section*{Acknowledgments}

This work was supported in part by NSF grant PHY-1912168. SB is supported in
part by the NSERC (funding reference \#CITA 490888-16) through a CITA National
Fellowship and by a McGill Space Institute fellowship. SC is supported by the
Sonata Bis Grant No. DEC-2017/26/E/ST2/00763 of the National Science Centre
Poland.

%\bibliographystyle{../preprint}
%\bibliography{../Bib/QuantGra,../Bib/Tunneling}

\end{document}